\documentclass[aps,prl,twocolumn,groupedaddress]{revtex4}

\usepackage[T1]{fontenc}
\usepackage[english]{babel}
\usepackage{graphicx}
\usepackage{color}
\usepackage{amsmath}
\usepackage{dsfont}
\usepackage{setspace}

\def\alf{\alpha}

\def\sign{\text{sign}}

\def\bp{{\bf p}}

\def\bQ{{\bf Q}}
\def\bS{{\bf S}}
\def\bn{{\bf n}}
\def\bR{{\bf R}}

\def\cA{{\cal A}}
\def\cE{{\cal E}}
\def\cH{{\cal H}}
\def\cD{{\cal D}}

\def\cG{{\cal G}}
\def\cP{{\cal P}}
\def\cU{{\cal U}}

\def\eps{\epsilon}

\def\im{\text{Im}\,}
\def\re{\text{Re}\,}

\def\sq#1{\sqrt{\vphantom{y^+} \smash{#1}}}
\def\sqr#1{\sqrt{\vphantom{\cD^{}_\bp}\smash{#1}}}

\def\sqwp{\sqrt{\vphantom{y^+} \smash{w^+_\bp}}}
\def\sqwm{\sqrt{\vphantom{y^+} \smash{w^-_\bp}}}

\begin{document}

\title{Non-Hermitian band topology from momentum-dependent relaxation in two-dimensional metals with spiral magnetism}

\author{Johannes~Mitscherling and Walter Metzner}
\affiliation{Max Planck Institute for Solid State Research, D-70569 Stuttgart, Germany}

\date{\today}

%
%
%
\begin{abstract}
We study the emergence of non-Hermitian band topology in a two-dimensional metal with planar spiral magnetism due to a momentum-dependent relaxation rate.
A sufficiently strong momentum dependence of the relaxation rate leads to exceptional points in the Brillouin zone, where the Hamiltonian is nondiagonalizable. The exceptional points appear in pairs with opposite topological charges and are connected by arc-shaped branch cuts.
We show that exceptional points inside hole and electron pockets, which are generally present in a spiral magnetic state with a small magnetic gap, can cause a drastic change of the Fermi surface topology by merging those pockets at isolated points in the Brillouin zone.
We derive simple rules for the evolution of the eigenstates under semiclassical motion through these crossing points, which yield geometric phases depending only on the Fermi surface topology.
The spectral function observed in photoemission exhibits Fermi arcs. Its momentum dependence is smooth -- despite of the nonanalyticities in the complex quasiparticle band structure.
\end{abstract}

\maketitle

%
%

\paragraph{Introduction.}

The discovery of topological insulators \cite{Hasan2010, Moore2010} has triggered a systematic analysis and classification of topological features of band structures in solids. So far, the main focus has been on noninteracting electrons and superconductors in a mean-field picture, corresponding to Hermitian quadratic Hamiltonians \cite{Chiu2016}. Recently, there has been growing interest in topological features of non-Hermitian Hamiltonians \cite{Bergholtz2021}. In quantum many-body systems, these naturally arise in certain open systems, but also as effective Hamiltonians capturing relaxation processes in interacting and/or disordered closed systems \cite{Kozii2017, Yoshida2018, Zyuzin2018, Kimura2019, McClarty2019, Moors2019, Papaj2019, Yoshida2019, Zyuzin2019, Aquino2020, Michishita2020, Nagai2020, Crippa2021, Michen2021, Rausch2021}. A striking effect, which is unique to non-Hermitian systems, is the existence of exceptional points in momentum space where the Hamiltonian is not diagonalizable. 

In this letter we show that the combination of two seemingly innocuous ingredients -- spiral magnetic order in a two-dimensional metal and a momentum-dependent relaxation rate -- can lead to a non-Hermitian Hamiltonian with non-trivial topological features, such as exceptional points and branch cuts in the Brillouin zone.
Spiral order is a candidate for incommensurate magnetic order observed in cuprate superconductors \cite{Shraiman1989, Machida1989, Dombre1990, Fresard1991, Chubukov1992, Chubukov1995, Kotov2004, Yamase2016, Eberlein2016, Mitscherling2018, Bonetti2020}, while relaxation rates with a pronounced momentum dependence arise naturally in two-dimensional systems with strong antiferromagnetic fluctuations \cite{Kampf1990, Vilk1997, Katanin2004, Rohe2005}.
We find that the momentum dependence of the relaxation rate can lead to a closing of the direct band gap between the quasi-particle bands $E^+_\bp$ and $E^-_\bp$ on one-dimensional branch cuts in the Brillouin zone, which terminate at exceptional points.
These lines of degenerate quasiparticle bands, which are sometimes referred to as non-Hermitian (bulk) Fermi arcs \cite{Kozii2017, Bergholtz2021}, are in general dispersive, in contrast to flat degenerate bands in some other systems \cite{Kozii2017, Nagai2020, Zhou2018}. In the dispersive case, hole and electron pockets merge at isolated momenta in the Brillouin zone where these degenerate bands cross the Fermi level, leading thus to a peculiar Fermi surface topology. Electrons traversing such crossing points along the Fermi surface acquire $\pi$-phase shifts, which can lead to a non-trivial geometric Berry phase.
Surprisingly, we find that the non-analyticity of the complex band at the exceptional points does not entail any singularity in the spectral function for single electron excitations. Nevertheless, the Fermi surface obtained from the spectral function seems truncated to Fermi arcs.


\paragraph{Spiral spin density waves.}

In a planar spiral spin density wave, the local magnetic moment has the form $\bS_i = m \bn_i$, where $m$ is a constant amplitude and $\bn_i$ a site-dependent unit vector, which rotates in a fixed but arbitrary plane. For definiteness, we choose $\bn_i$ to lie in the $x$-$y$ plane, such that $\bn_i = \big(\cos(\bQ\cdot\bR_i),\sin(\bQ\cdot\bR_i),0\big)$, where $\bQ$ is the wave vector of the spin density wave. On a mean-field level, the planar spiral spin density wave is described by the two-band tight-binding Hamiltonian $H = \sum_\bp \Psi^\dag_\bp H^{}_\bp \Psi^{}_\bp$, where the spinor $\Psi_\bp = \left( c_{\bp+\bQ,\uparrow}, c_{\bp,\downarrow} \right)$ collects the two spin components with a relative momentum shift $\bQ$, and
\begin{align}
 H_\bp =
 \begin{pmatrix}
 \eps_{\bp+\bQ} & -\Delta \\[2mm] -\Delta & \eps_\bp
 \end{pmatrix} \, ,
 \label{eqn:DefHBare}
\end{align}
where $\eps_\bp$ is the (bare) band dispersion and $\Delta$ the magnetic gap \cite{Igoshev2010}. The wave vector $\bQ$ can be chosen arbitrarily. The simple two-band structure is due to a symmetry under combined lattice translations and spin rotations \cite{Sandratskii1998}. For $\bQ = (0,0)$ and $\bQ = (\pi,\pi)$ one recovers ferromagnetic and N\'eel antiferromagnetic order aligned in the $xy$ plane, respectively. 
In the following we consider incommensurate spiral order with wave vectors of the form $\bQ = (\pi-2\pi\eta,\pi)$ with $\eta>0$, as found in the hole-doped Hubbard and $t$-$J$ model \cite{Shraiman1989, Machida1989, Dombre1990, Fresard1991, Chubukov1992, Chubukov1995, Kotov2004,Yamase2016,Eberlein2016,Mitscherling2018, Bonetti2020}.
Diagonalizing $H_\bp$ one finds two quasi-particle bands $E^\pm_\bp$ with a minimal direct gap $\Delta$, which generally results in a reconstructed Fermi surface with electron and/or hole pockets \cite{Eberlein2016, Mitscherling2018, Bonetti2020}.


\paragraph{Non-Hermitian effective Hamiltonian.}

In interacting electron systems all the information on the Fermi surface, quasi-particle bands and decay rates, as well as the spectral function measured in photoemission, is encoded in the single-particle Green's function. The bare Green's function of the noninteracting reference system is dressed by the self-energy, which receives contributions from the electron-electron interaction, and possibly from phonon and impurity scattering. In the low-frequency limit, the real part of the self-energy yields a renormalization of the band structure, and a reduction of the quasiparticle weight, while the imaginary part describes the quasiparticle relaxation rate $\Gamma_\bp$. Here we discard the real part and focus on the more interesting effects of
$\Gamma_\bp$ in combination with spiral magnetic order. In the two-component spinor basis defined above, the retarded Green's function can then be written as
\begin{align}
 \cG^R_\bp(\omega)=\big[\omega+\mu-\cH_\bp\big]^{-1}\, ,
\end{align}
where the {\it non-Hermitian}\/ Hamiltonian $\cH_\bp$ is defined as
\begin{align}
\cH_\bp & = H_\bp-
i\begin{pmatrix}
  \Gamma_{\bp+\bQ} & 0 \\[2mm]
  0 & \Gamma_\bp
 \end{pmatrix} \, ,
 \label{eqn:DefH}
\end{align}
with $H_\bp$ from Eq.~(\ref{eqn:DefHBare}).
Imaginary off-diagonal components have only minor consequences \cite{suppl}.


\paragraph{Exceptional points.} 

$\cH_\bp$ has the complex eigenvalues
\begin{align}
 \cE^\pm_\bp = \eps^s_\bp \pm \sqrt{\cD_\bp} - i\Gamma^s_\bp \, ,
 \label{eqn:Epm}
\end{align}
with the discriminant
\begin{align}
 \cD^{}_\bp = (\eps^a_\bp - i\Gamma^a_\bp)^2 + \Delta^2 \, ,
 \label{eqn:disc}
\end{align}
where $\eps^{s/a}_\bp = \frac{1}{2}(\eps_{\bp+\bQ} \pm \eps_\bp)$ and $\Gamma^{s/a}_\bp = \frac{1}{2}(\Gamma_{\bp+\bQ} \pm \Gamma_\bp)$ are symmetric and antisymmetric linear combinations.
The condition $\cD_\bp = 0$ defines the exceptional points, that is, the set of momenta at which $\cH_\bp$ is not diagonalizable. The momentum dependence of $\Gamma_\bp$ is crucial for the existence of exceptional points. For $\Gamma_\bp = \Gamma$ or $\Gamma_\bp = \Gamma_{\bp+\bQ}$, we have $\Gamma^a_\bp = 0$, and thus $\cD_\bp \geq \Delta^2 > 0$ in the magnetically ordered phase. For $\Gamma^a_\bp \neq 0$, the real and imaginary parts of $\cD_\bp$ yield two conditions for an exceptional point, 
\begin{align}
 & \eps_\bp = \eps_{\bp+\bQ} \, , \label{eqn:Cond1} \\[2mm]
 & |\Gamma_{\bp+\bQ} - \Gamma_\bp| = 2\Delta \label{eqn:Cond2}\, ,
\end{align}
which need to be satisfied simultaneously.
The second condition requires a relaxation rate that exceeds the magnetic gap at particular momenta. The exceptional points can be classified by a topological charge $\nu_i = \pm \frac{1}{2}$ via
\begin{align}
 \nu_i = - \frac{1}{2\pi}\oint_{\Gamma_i} d\bp \cdot \nabla^{}_\bp 
 \arg\big[\cE^+_\bp - \cE^-_\bp\big] \, ,
 \label{eqn:vorticity}
\end{align}
where $\Gamma_i$ is a closed contour encircling the $i$-th exceptional point counterclockwise \cite{Kozii2017, Shen2018, Yang2021}.
Exceptional points with opposite charge are connected by branch cuts where $\cD_\bp$ is real and negative.


\paragraph{Fermi surface reconstruction.}

As an example, we assume a tight-binding dispersion on a square lattice, a magnetic gap $\Delta$ and a wave vector $\bQ=(\pi-2\pi\eta,\pi)$ such that several hole and electron pockets are present, confined by momenta at which the lower and upper quasiparticle bands cross the Fermi level, respectively. The dispersion has the form
\begin{align}
 \eps_\bp =& - 2t(\cos p_x + \cos p_y) - 4t' \cos p_x \cos p_y \nonumber \\
 & - 2t''[\cos(2p_x) + \cos(2p_y)] ,
 \label{eqn:dispersion}
\end{align}
where $t$, $t'$, and $t''$ are hopping amplitudes between nearest, next-nearest, and third-nearest neighbors, respectively.
We use $t$ as our energy unit, and we choose $t'/t = -0.17$ and $t''/t = 0.05$, as widely used for $\text{La}_{2-x}\text{Sr}_x\text{CuO}_4$ (LSCO) superconductors \cite{Bonetti2020, Mitscherling2018, Verret2017}. 
The parameters of the magnetic order $\Delta/t = 0.144$ and $\eta=0.106$ are taken from recent DMFT results for the two-dimensional Hubbard model with LSCO parameters at a hole doping $p = 1-n = 0.177$ \cite{Bonetti2020}.

For the momentum-dependent relaxation rate $\Gamma_\bp$ we assume a $d$-wave form
\begin{align}
 \Gamma_\bp = \gamma_0 + \frac{\gamma_d}{4}(\cos p_x-\cos p_y)^2 \, ,
 \label{eqn:gamma}
\end{align}
with $\gamma_0, \gamma_d \geq 0$, which has its minimal value ($\Gamma_{\rm min} = \gamma_0$) along the Brillouin zone diagonals, and its maximal value ($\Gamma_{\rm max} = \gamma_0 + \gamma_d$) at the points $(\pi,0)$ and $(0,\pi)$ on the $p_x$ and $p_y$ axis.
The relaxation rate in a cuprate compound from the LSCO family has recently been determined experimentally via angle-resolved magneto-resistance measurements in the overdoped regime at various temperatures \cite{Fang2020}, yielding an estimate $\gamma_0/t \approx 0.015$ and $\gamma_d/t \approx 0.15$.
For a given relaxation rate $\Gamma_\bp$, the condition \eqref{eqn:Cond2} can always be satisfied for a sufficiently small gap, for instance, near the onset of magnetic order at a quantum critical point.
For a better visualization of the topological effects, we choose a sizable magnetic gap and consider relatively large values for the relaxation rate, namely $\gamma_0/t = 0.05$, and $\gamma_d/t \leq 1.6$.
We fix the doping level at $p = 0.177$. All results are obtained at zero temperature.
\begin{figure*}[t!]
\centering
   \includegraphics[width=0.24\textwidth]{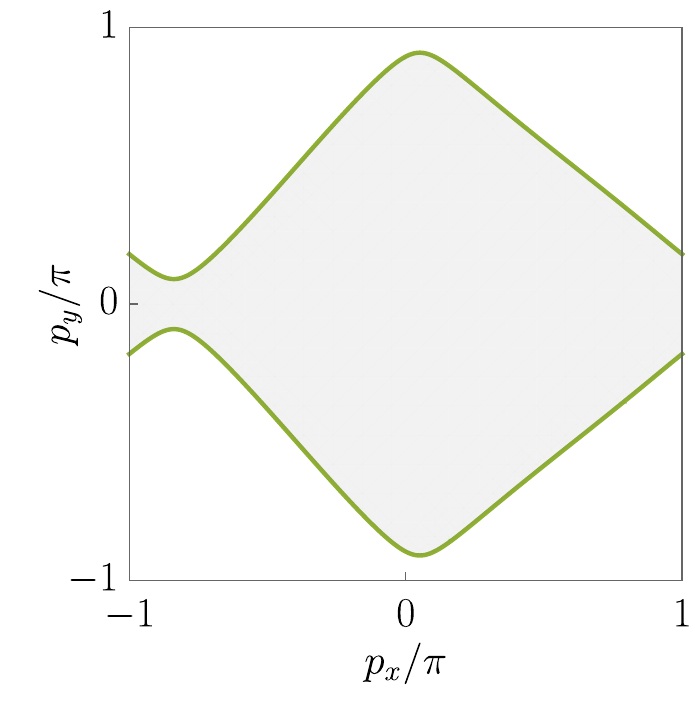}
   \includegraphics[width=0.24\textwidth]{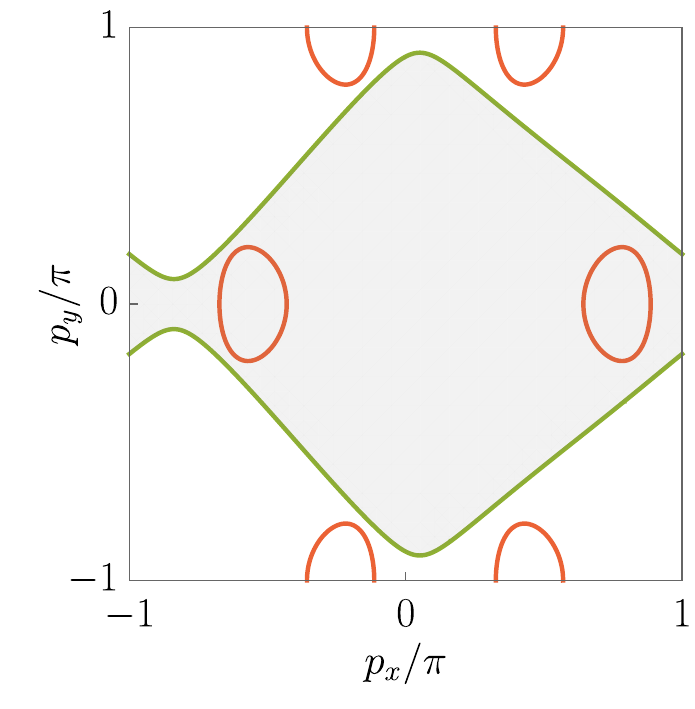}
   \includegraphics[width=0.24\textwidth]{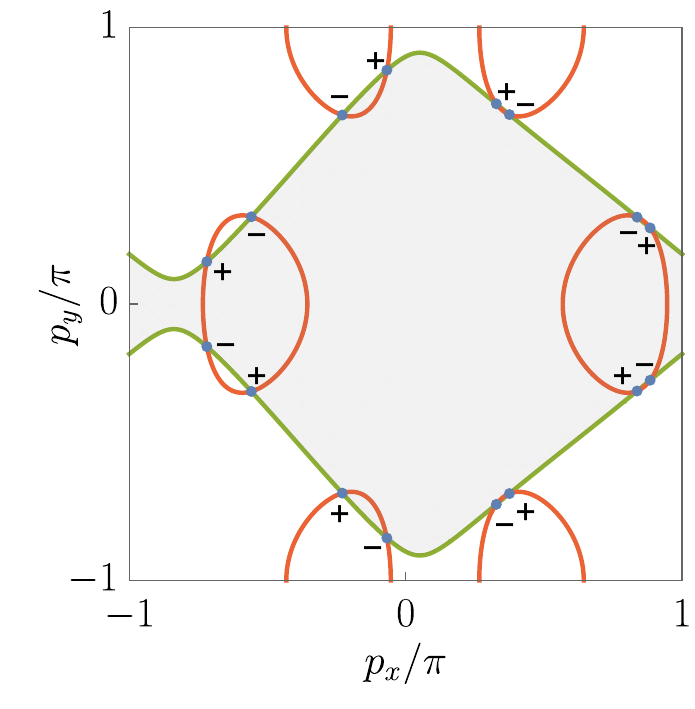}
   \includegraphics[width=0.24\textwidth]{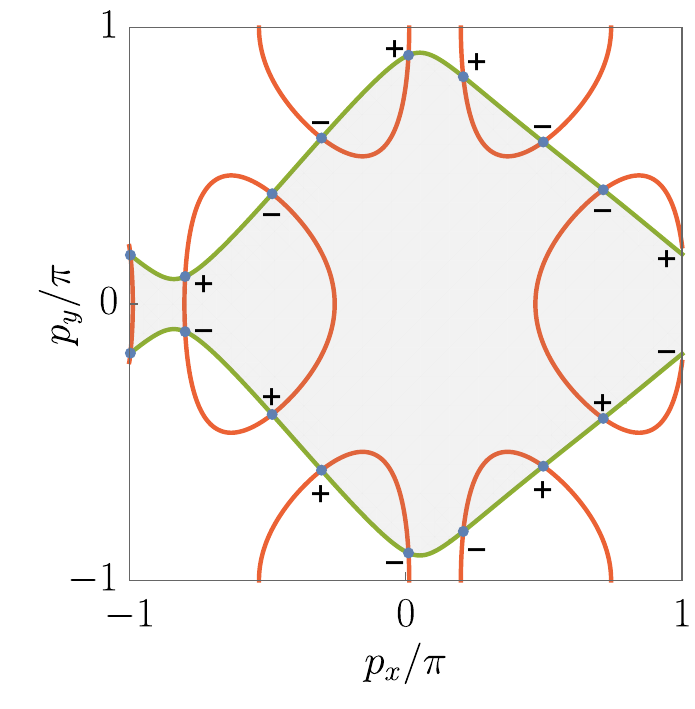}\\
   \includegraphics[width=0.24\textwidth]{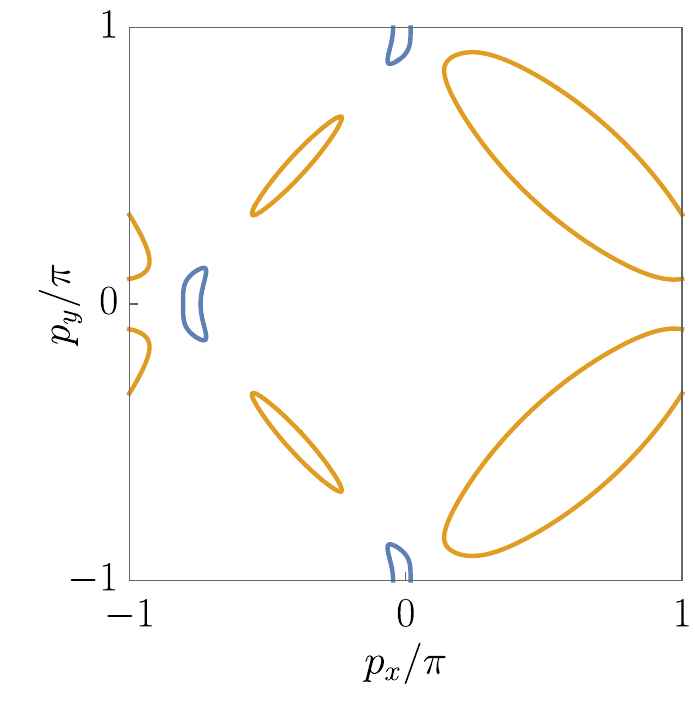}
   \includegraphics[width=0.24\textwidth]{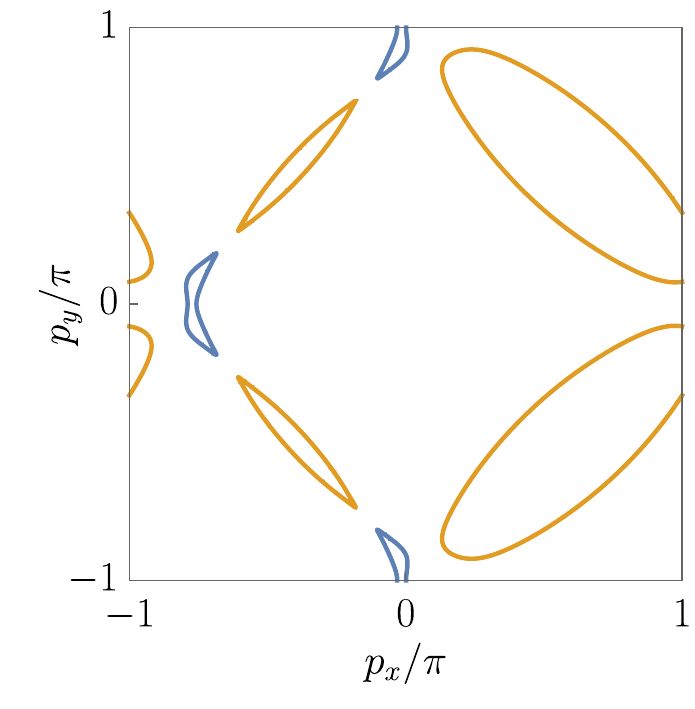}
   \includegraphics[width=0.24\textwidth]{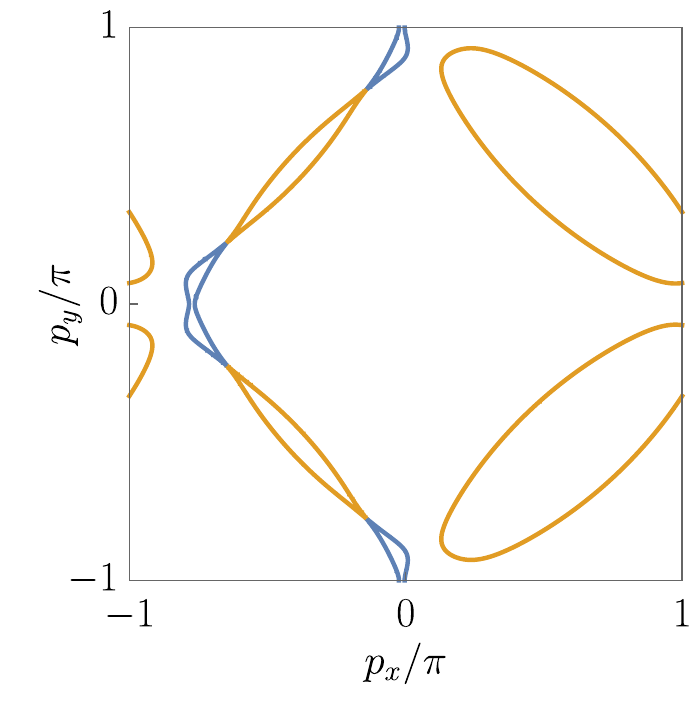}
   \includegraphics[width=0.24\textwidth]{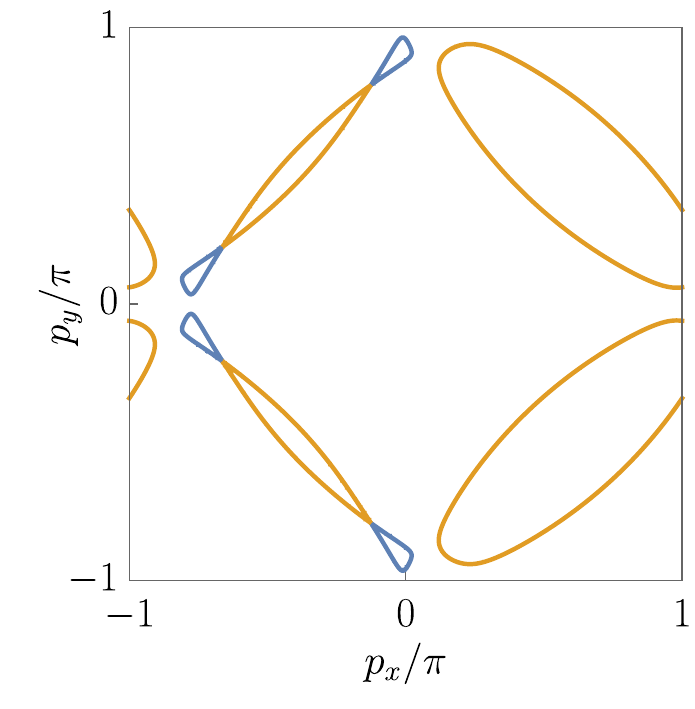}
\caption{{\it From left to right:} $\gamma_d/t=0$, $0.8$, $1.0$, and $1.6$.
{\it Upper row:} The conditions given by Eq.~\eqref{eqn:Cond1} (green ``nesting'' lines), which separates the region of $\eps^a_\bp>0$ (gray) and $\eps^a_\bp<0$ (white), and Eq.~\eqref{eqn:Cond2} (red lines). Exceptional points are situated at the intersection of both lines, and carry the topological charge $\nu_i = \pm \frac{1}{2}$. {\it Lower row:} The quasiparticle Fermi surfaces defined by $\re E^\pm_\bp = \mu$ for fixed doping p=0.177. Hole pockets (orange) and electron pockets (blue) merge at isolated momenta on the branch cuts, that is, on the parts of the nesting line between exceptional points of opposite charge. 
\label{fig:FermiSurface}}
\end{figure*}

In Fig.~\ref{fig:FermiSurface} (top row) we show 
the ``nesting'' lines defined by Eq.~\eqref{eqn:Cond1}, and the lines corresponding to the condition~\eqref{eqn:Cond2} 
for different $\gamma_d/t$.
Exceptional points where these lines cross exist for $\gamma_d/t \geq 1.0$. Changing parameters, exceptional points can be created or annihilated only in pairs with opposite topological charge.
In the bottom row of Fig.~\ref{fig:FermiSurface} we show the quasiparticle Fermi surfaces. Electron and hole pockets are disconnected for $\gamma_d/t = 0$ and $0.8$, while for larger $\gamma_d/t$ they merge at isolated momenta on the branch cuts. 
\begin{figure}[t!]
\centering
   \includegraphics[width=0.232\textwidth]{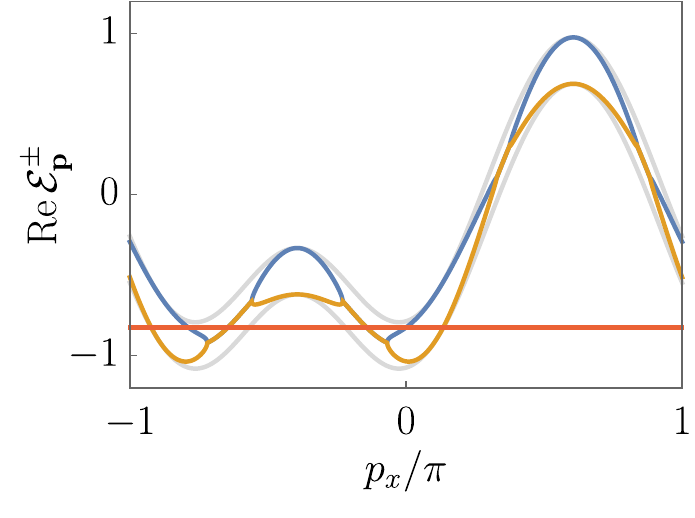}
   \hspace{1mm}
   \includegraphics[width=0.232\textwidth]{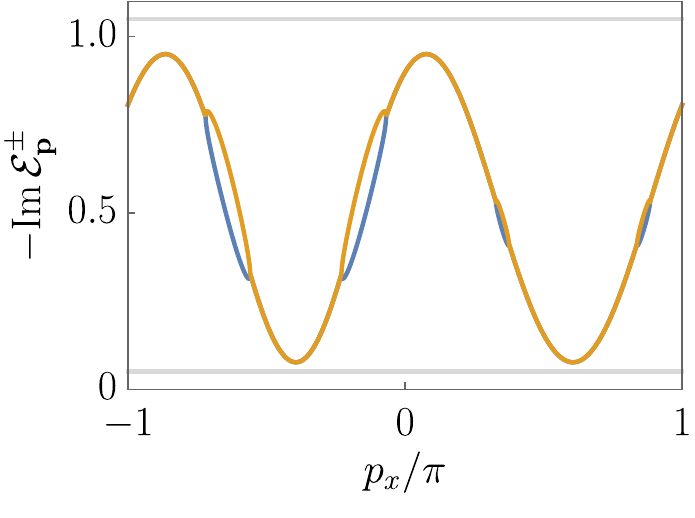}
\caption{Real part (left) and imaginary part (right) of the complex quasiparticle bands
 $\cE^\pm_\bp$ for $\gamma_d/t=1$ on the upper nesting line 
 where
 $\eps_\bp = \eps_{\bp+\bQ}$
 (see upper row in Fig.~\ref{fig:FermiSurface}) as a function of $p_x$.
 {\it Left:} The band gap closes between the exceptional points. The bands for $\gamma_d=0$
 are shown for comparison (gray lines). The chemical potential $\mu$ is indicated by the red line.
 {\it Right:} The quasiparticle relaxation rate $\Gamma^\pm_\bp = -\im\cE^\pm_\bp$ is
 double-valued between the exceptional points. The gray lines indicate $\Gamma_\text{min}$
 and $\Gamma_\text{max}$. Note that $\Gamma^\pm_\bp$ is always positive.\label{fig:Bands}}
\end{figure}

In Fig.~\ref{fig:Bands} we show the real and imaginary parts of the quasiparticle bands $\cE^\pm_\bp$ for $\gamma_d/t=1$ as a function of $p_x$ along the upper nesting line in Fig.~\ref{fig:FermiSurface}, where the discriminant $\cD_\bp$ is real (since $\eps_\bp^a = 0$). For $|\Gamma_{\bp+\bQ} - \Gamma_\bp| < 2\Delta$, the square root in Eq.~\eqref{eqn:Epm} is real, such that $E_\bp^\pm = \re\cE^\pm_\bp$ describes two separate bands. For $|\Gamma_{\bp+\bQ} - \Gamma_\bp| > 2\Delta$, that is, on the branch cut, the square root in Eq.~\eqref{eqn:Epm} is purely imaginary such that $E_\bp^+ = E_\bp^-$, while now $\Gamma_\bp^\pm = - \im\cE_\bp^\pm$ assumes two distinct values. At the exceptional points, where $|\Gamma_{\bp+\bQ} - \Gamma_\bp| = 2\Delta$, both real and imaginary parts of the two complex bands $\cE^\pm_\bp$ collapse to a single value.
The degenerate band $E^+_\bp = E^-_\bp$ on the branch cut is dispersive. Thus, it intersects the Fermi level only at isolated momenta, which leads to the peculiar Fermi surface topology in Fig.~\ref{fig:FermiSurface}. 
The merging of hole and electron pockets at single isolated momenta is a generic consequence of exceptional points with opposite topological charge inside the pockets and, thus, not restricted to our specific realization by the particular form of the dispersion in Eq.~\eqref{eqn:dispersion} or the relaxation rate in Eq.~\eqref{eqn:gamma}.


\paragraph{Semiclassical transport through crossing points.}

\begin{figure}[t!]
\centering
\includegraphics[width=0.30\textwidth]{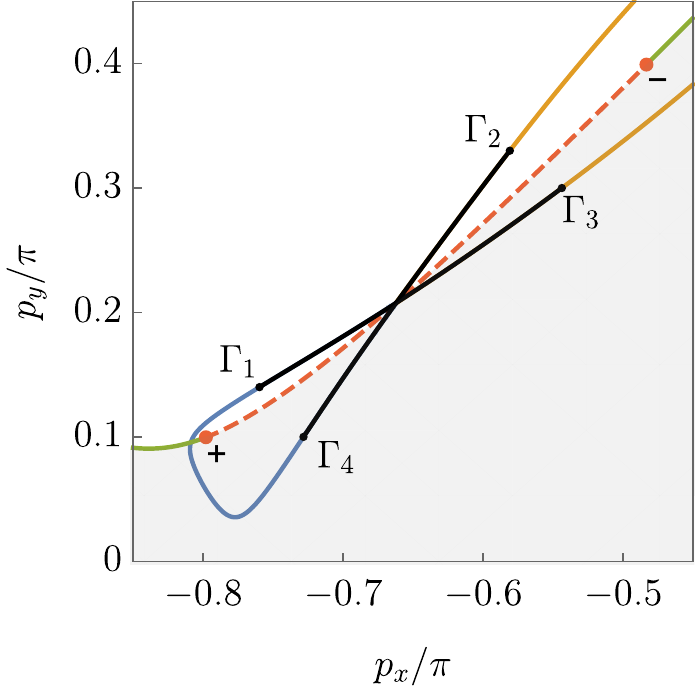}
\caption{Close-up of a crossing point for $\gamma_d/t = 1.6$. The electron (blue) and hole (orange) pockets encircle exceptional points (red dots) of opposite charge. They merge at one point on the branch cut (red-dashed line). Only the two diagonal paths from $\Gamma_1$ to $\Gamma_3$ and from $\Gamma_2$ to $\Gamma_4$ are continuously connected. The crossing rules depend on the sign change between the regions $\eps^a_\bp > 0$ (gray) and $\eps^a_\bp < 0$ (white). \label{fig:Crossing}}
\end{figure}

In a semiclassical description of the electron dynamics, the momentum of electrons changes smoothly in the direction of the applied force \cite{Ashcroft1976}. The Lorentz force acts perpendicularly to the electron velocity, such that a magnetic field makes low-energy electrons move along the Fermi surface. We now clarify how electrons move semiclassically through the crossing points.
There are potentially six paths on the Fermi surface.
(see Fig.~\ref{fig:Crossing}). We study the evolution of a biorthonormal basis with left and right eigenstates $|L^n_\bp\rangle$ and $|R^n_\bp\rangle$ for the bands $n=\pm$ when $\bp$ traverses the crossing point \cite{Brody2014, footnote_RL}.
Since the Hamiltonian in Eq.~\eqref{eqn:DefH} is symmetric, $\cH^{}_\bp=\cH_\bp^\text{tr}$, we can choose a gauge such that $|L^n_\bp\rangle = \big(|R^n_\bp\rangle\big)^*$. Thus, the Berry connection 
$i\langle L^\pm_\bp|\partial^{}_{p_\alf} R^\pm_\bp\rangle$ vanishes and the geometric phase $\gamma_B$ is determined exclusively by the overlap of the initial and final states \cite{Keck2003}. For definiteness, the remaining gauge freedom $|R^n_\bp \rangle \rightarrow \pm|R^n_\bp \rangle$ has also been fixed.

In the Supplemental Material \cite{suppl} we show that only the two diagonal paths allow for a continuous evolution of the eigenstates through the crossing point. The phase shift is determined by the sign change of $\eps^a_\bp$. With the shorthand notation $|n_\bp\rangle = |R^n_\bp\rangle$, the transition of states at the crossing point is given by
\begin{align}
 |+_\bp\rangle \rightleftharpoons -|-_\bp\rangle \quad \mbox{and} \quad
 |-_\bp\rangle \rightleftharpoons |+_\bp\rangle \, ,
 \label{eqn:Evolution}
\end{align}
where the transition from left to right is when crossing with sign change $+\rightarrow -$ (gray to white in Fig.~\ref{fig:Crossing}) and the evolution from right to left is when crossing with sign change $- \rightarrow +$ (white to gray in Fig.~\ref{fig:Crossing}). The first rule in Eq.~\eqref{eqn:Evolution} involves a minus sign beside the well-known swapping of the eigenstates \cite{Bergholtz2021}.
The velocities $\partial_{p_\alf}E^\pm_\bp$ are smooth and finite at the crossing point \cite{suppl}. Note that the rules in Eq.~\eqref{eqn:Evolution} are gauge dependent, but the total geometric phase accumulated in a closed loop is a gauge independent quantity.

\begin{figure}[t!]
\centering
   \includegraphics[width=0.23\textwidth]{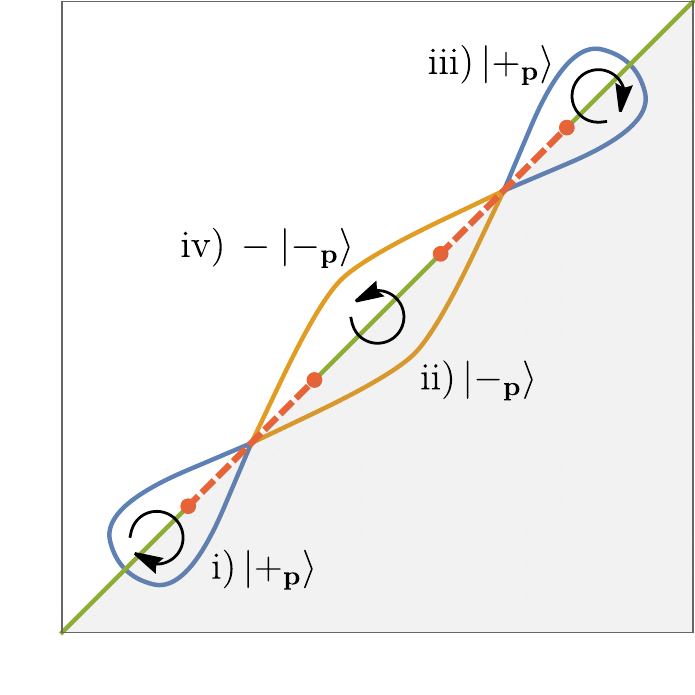}
   \includegraphics[width=0.23\textwidth]{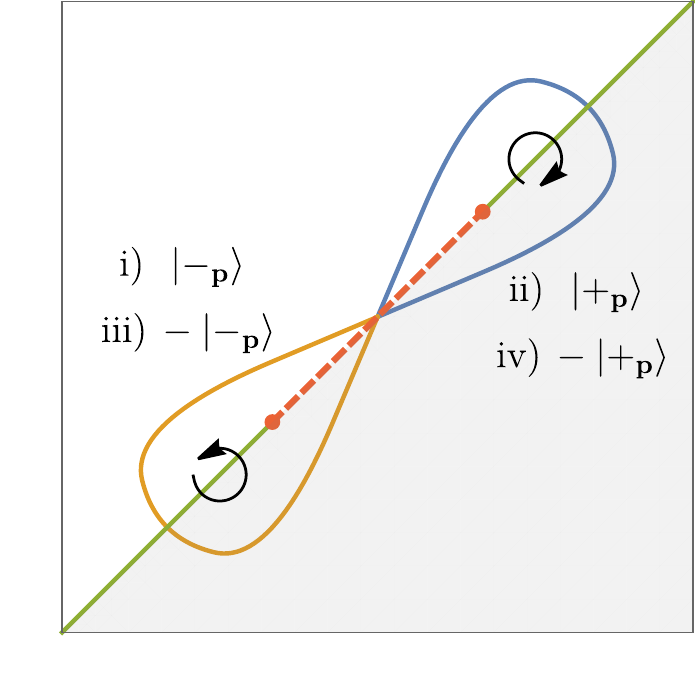}
\caption{Evolution of the 
eigenstates $\text{i)}-\text{iv)}$ for electrons moving along the Fermi surface in the arrow directions.
{\it Left:} The ``double-eight'' topology appearing in Fig.~\ref{fig:FermiSurface}.
{\it Right:} A hypothetical ``eight'' topology.
\label{fig:AdiabaticEvolution}}
\end{figure}

In Fig.~\ref{fig:AdiabaticEvolution} we sketch the evolution of the eigenstates for electrons moving along the Fermi surface according to Eq.~\eqref{eqn:Evolution}. For the pockets in Fig.~\ref{fig:FermiSurface} we find a vanishing geometric phase $\gamma_B = 0$ after a completed round, but a relative phase difference $\pi$ on opposite sides of the hole pocket [step ii) and iv)]. We predict a geometric phase $\gamma_B = \pi$ for an ``eight'' topology of a merged electron and hole pocket, so that the original state is then recovered only after two rounds.

Quantum oscillation experiments at sufficiently large magnetic fields $\omega_c \tau > 1$, where $\omega_c$ is the cyclotron frequency and $\tau = 1/2\Gamma$, can be used to measure the Fermi surface topology. Thus, the merging of electron and hole pockets (see Fig.~\ref{fig:FermiSurface}) is visible at least in principle in the spectrum of quantum oscillations. 
A geometric phase can be observed experimentally as a phase shift in quantum oscillations \cite{Shoenberg1984}. A detailed microscopic or semiclassical analysis of transport in non-Hermitian systems is still ongoing research \cite{Mitscherling2018, Xu2017, Chen2018, Philip2018, Hirsbrunner2019, Wang2019, Mitscherling2020, Silberstein2020, Groenendijk2021, Wang2021} and beyond the scope of this paper.
%


\paragraph{Spectral functions.}

The quasiparticle spectral function is given by the diagonal matrix $\tilde \cA_\bp(\omega) = -\frac{1}{\pi}\im \tilde \cG^R_\bp(\omega)$, where $\tilde \cG^R_\bp(\omega) = [\omega+\mu-\tilde \cH_\bp]^{-1}$ is the retarded Green's function in the quasiparticle basis, and $\tilde \cH_\bp$ is the diagonalized non-Hermitian Hamiltonian with the eigenvalues $\cE^\pm_\bp$ from Eq.~\eqref{eqn:Epm}.
The quasiparticle spectral functions are thus Lorentzians with positions $\eps^s_\bp \pm \eps^\cD_\bp-\mu$ and widths $\Gamma^s_\bp \mp \Gamma^\cD_\bp$, where $\eps^\cD_\bp = \re \sqrt{\cD_\bp}$ and $\Gamma^\cD_\bp = \im \sqrt{\cD_\bp}$.

The spectral function matrix in the bare band basis is given by
$\cA_\bp(\omega) = \frac{1}{2\pi i} \big( \cG^A_\bp(\omega) - \cG^R_\bp(\omega)\big)$ with $\cG^A_\bp(\omega) = [\cG^R_\bp(\omega)]^\dag$.
The Green's function in the bare band basis is related to the quasiparticle Green's function by $\cG^R_\bp(\omega) = \cU^{}_\bp \, \tilde\cG^R_\bp(\omega) \, \cU^{-1}_\bp$, where the matrix $\cU^{}_\bp$ diagonalizes $\cH_\bp$
for all momenta except, of course, the exceptional ones.
The diagonal elements of $\cA_\bp(\omega)$ are obtained as \cite{suppl}
\begin{alignat}{2}
 &A^{\uparrow/\downarrow}_\bp(\omega) &&=
 \frac{1}{2}\Big[\tilde A^+_\bp(\omega) + \tilde A^-_\bp(\omega)\Big] \label{eqn:A1}\\[3mm]
 & && \pm \bigg( \frac{1}{2}\frac{\eps^a_\bp \eps^\cD_\bp - \Gamma^a_\bp\Gamma^\cD_\bp}
 {(\eps^\cD_\bp)^2+(\Gamma^\cD_\bp)^2} 
 \Big[\tilde A^+_\bp(\omega) - \tilde A^-_\bp(\omega)\Big] 
 \label{eqn:A2}\\[2mm]
 & && + \frac{1}{2\pi}\frac{\eps^a_\bp \Gamma^\cD_\bp + \Gamma^a_\bp\eps^\cD_\bp}
 {(\eps^\cD_\bp)^2 + (\Gamma^\cD_\bp)^2} 
 \Big[\tilde P^+_\bp(\omega) - \tilde P^-_\bp(\omega)\Big] \bigg) \, , \label{eqn:A3}
\end{alignat}
where $\tilde P^\pm_\bp(\omega)$ are the elements of the diagonal matrix  $\tilde\cP_\bp(\omega) = \re\tilde\cG_\bp^R(\omega)$. Due to the momentum shift $\bQ$ in the spinor $\Psi_\bp$, the total spectral function for the physical single electron excitations reads
\begin{align}
 A^{}_\bp(\omega) = A^\uparrow_{\bp-\bQ}(\omega) + A^\downarrow_\bp(\omega) \, .
 \label{eqn:spectralfunction}
\end{align}
For $\Gamma^a_\bp = 0$ we have $\Gamma^\cD_\bp = 0$, so that we recover the well-known result for a momentum-independent relaxation rate \cite{suppl,Eberlein2016}. The appearance of the term \eqref{eqn:A3} is directly linked to a nonzero $\Gamma^a_\bp$. In the Supplemental Material \cite{suppl} we analyze the effect of exceptional points. Both the second term in \eqref{eqn:A2} and the third term in \eqref{eqn:A3} are discontinuous at the exceptional points. Due to the phase shift $\frac{\pi}{2}$ in $\sqrt{\cD_\bp}$ when crossing the exceptional point, the two contributions are mapped onto each other. Thus, the sum of both is continuous. In other words, the nonanalyticity of the complex band at the exceptional points does not appear in $A^{}_\bp(\omega)$.

\begin{figure}[t!]
\centering
 \includegraphics[width=0.23\textwidth]{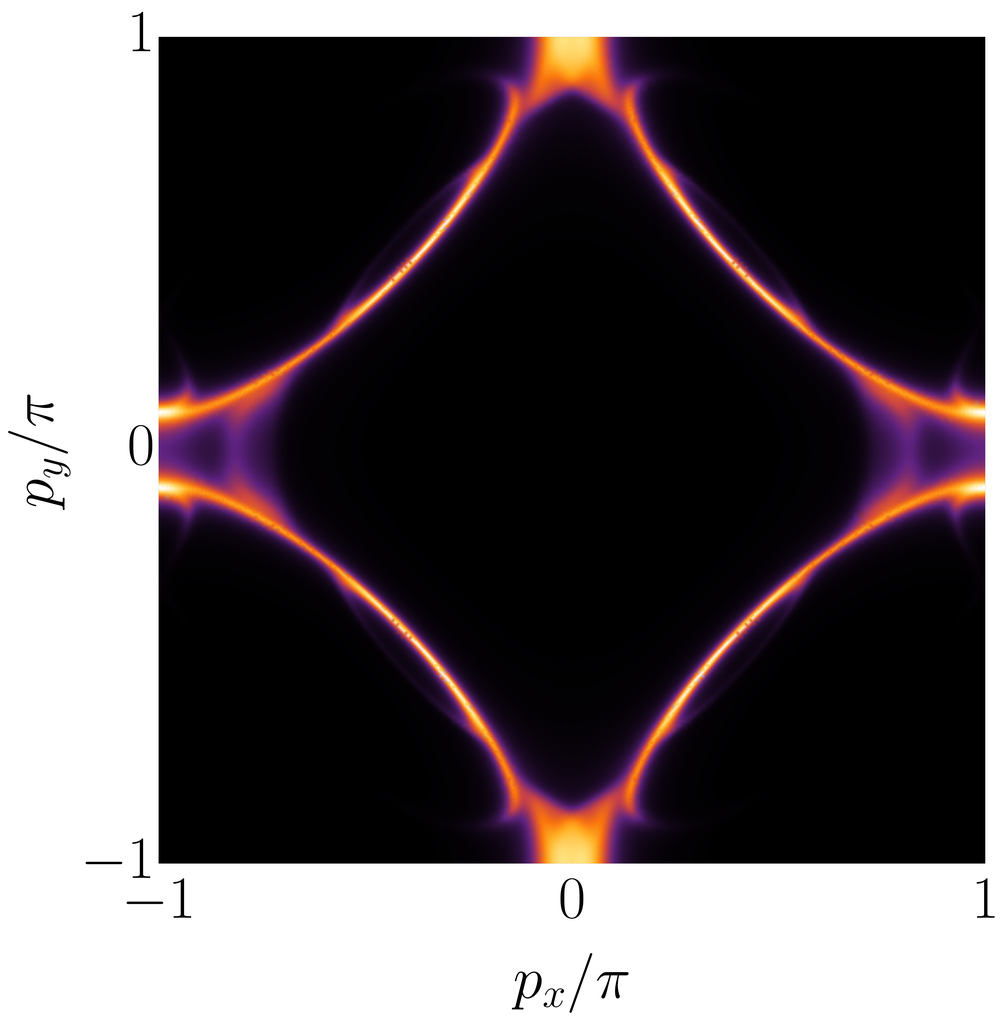}
 \hspace{1mm}
 \includegraphics[width=0.23\textwidth]{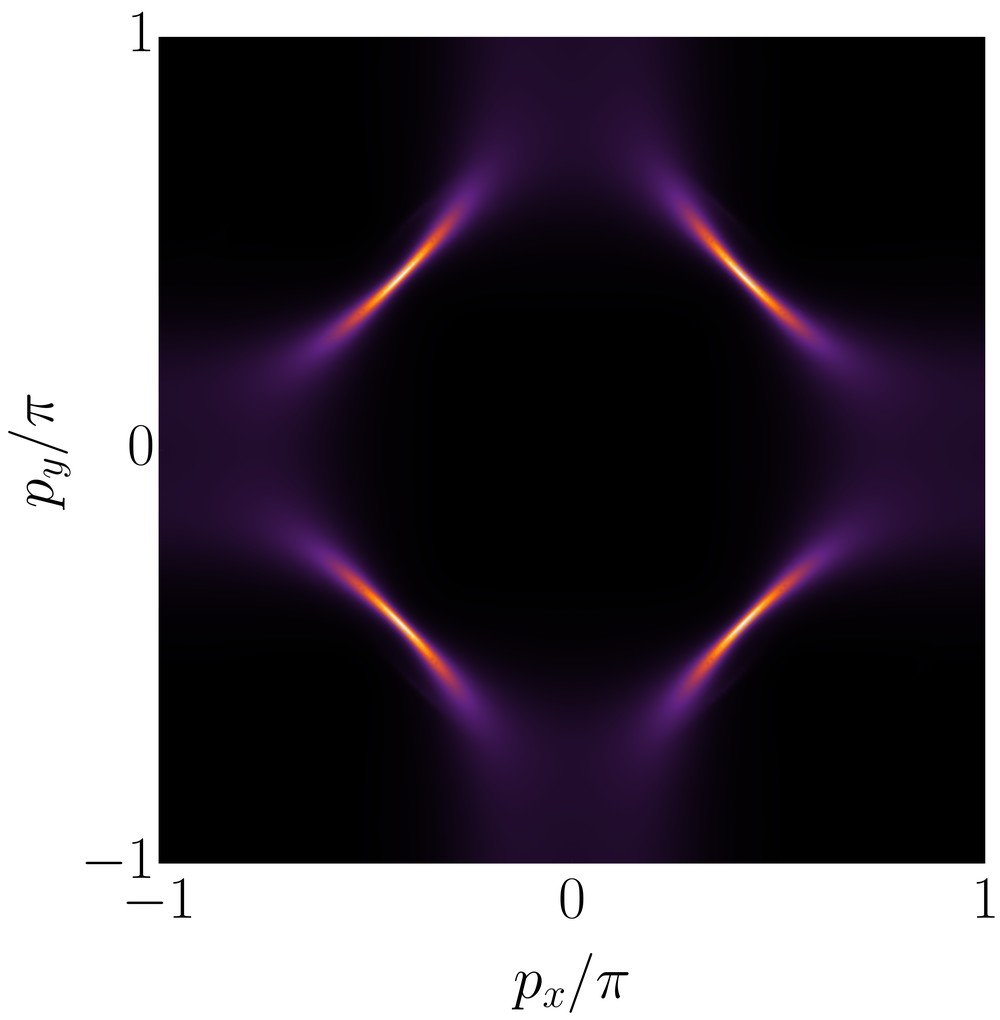}
\caption{The spectral function for single particle excitations $A_\bp(0)$
 for $\gamma_d/t=0$ (left) and $\gamma_d/t=1$ (right). \label{fig:ASpiral}}
\end{figure}

In Fig.~\ref{fig:ASpiral} we show the spectral function $A_\bp(\omega)$ at $\omega = 0$ for $\gamma_d/t =0$ and $\gamma_d/t = 1$. The spectral weight is strongly suppressed for momenta away from the bare Fermi surface \cite{suppl, Eberlein2016}. Moreover, the angle dependence of $\Gamma_\bp$ reduces the spectral weight in the antinodal region, such that only Fermi arcs in the nodal region are visible.
%


\paragraph{Conclusions.}

We have analyzed the non-Hermitian band topology resulting from a momentum-dependent relaxation rate $\Gamma_\bp$ in a two-dimensional metal with spiral magnetic order.
We provided a concrete example for a specific band dispersion and relaxation rate.
We find that
arc-shaped branch cuts connecting exceptional points with opposite topological charges appear in the Brillouin zone.
Exceptional points inside hole and electron pockets lead to a peculiar Fermi surface topology with pockets merging at isolated points in the Brillouin zone.
We have derived rules for the evolution of eigenstates under semiclassical motion through these crossing points, from which geometric phases associated with the Fermi surface topology can be obtained. The change of the Fermi surface topology and the geometric phase are visible at least in principle via quantum oscillations.
The spectral function for single-particle excitations, which can be observed in photoemission experiments, exhibits Fermi arcs. Its momentum dependence is however smooth, due to subtle cancellations of the non-analyticities in the complex quasiparticle band structure.

Our work provides an example for an intriguing non-Hermitian topological band structure emerging from a combination of conventional ingredients, in an electron system that was hitherto expected to be topologically trivial. Following this paradigm, we expect the discovery of other condensed matter systems with an interesting non-Hermitian band topology.


\paragraph{Acknowledgments.} We are very grateful to P.~M.~Bonetti, G.~Grissonnanche, S.~Groenendijk, M.~M.~Hirschmann, and especially A.~Schnyder for valuable discussions.



\onecolumngrid
\clearpage
\begin{center}
\textbf{\large Supplemental Material for ``Non-Hermitian band topology from momentum-dependent relaxation in two-dimensional metals with spiral magnetism''}
\end{center}

\setcounter{equation}{0}
\setcounter{figure}{0}
\setcounter{table}{0}
\setcounter{page}{1}
\makeatletter
\renewcommand{\theequation}{S\arabic{equation}}
\renewcommand{\thefigure}{S\arabic{figure}}
\renewcommand{\bibnumfmt}[1]{[S#1]}
\renewcommand{\citenumfont}[1]{S#1}

\spacing{1.25}


In this Supplemental Material we give further details on (i) the special case of a non-Hermitian part proportional to the identity matrix, (ii) the consequences of off-diagonal self-energy components, (iii) the behavior of the quasiparticle bands and eigenstates when crossing the branch cut, and (iv) the derivation and properties of the spectral functions.

\section{Special case of a non-Hermitian part proportional to the identity matrix}

In this section we briefly discuss the special case where the non-Hermitian part in Eq.~\eqref{eqn:DefH} is proportional to the identity matrix, which applies to a constant relaxation rate $\Gamma_\bp = \Gamma$ and, more generally, to a momentum-dependent relaxation rate obeying $\Gamma_\bp = \Gamma_{\bp+\bQ}$. The latter case is fulfilled for a ferromagnetic wave vector $\bQ=(0,0)$ and for a N\'eel antiferromagnetic wave vector $\bQ=(\pi,\pi)$ with a scattering rate of the $d$-wave form in Eq.~\eqref{eqn:gamma}. The (non-Hermitian) Hamiltonian reads 
\begin{align}
\cH_\bp =
\begin{pmatrix}
 \eps^s_\bp+\eps^a_\bp & -\Delta \\[2mm]
 -\Delta & \eps^s_\bp-\eps^a_\bp
\end{pmatrix}-
i\begin{pmatrix}
  \Gamma_\bp & 0 \\[2mm]
  0 & \Gamma_\bp
 \end{pmatrix} \, ,
 \label{eqn:H_diag}
\end{align}
where the first term is the Hermitian Hamiltonian $H_\bp$ in Eq.~\eqref{eqn:DefHBare}. Since the non-Hermitian part is proportional to the identity matrix, it is still possible to diagonalize $\cH_\bp$ by the same unitary transformation that diagonalizes the Hermitian matrix $H_\bp$. The eigenvectors of the upper and lower band are
\begin{align}
 |+_\bp\rangle=\frac{1}{\sqwp}
 \begin{pmatrix}
  \eps^a_\bp+\sqrt{(\eps^a_\bp)^2+\Delta^2} \\[3mm] -\Delta
 \end{pmatrix}\, ,
 \hspace{1cm}
 |-_\bp\rangle=\frac{1}{\sqwm}
 \begin{pmatrix}
  \eps^a_\bp-\sqrt{(\eps^a_\bp)^2+\Delta^2} \\[3mm] -\Delta
 \end{pmatrix} \, .
 \label{eqn:eigen_diag}
\end{align}
with the normalization $w^\pm_\bp=\Delta^2+\big(\eps^a_\bp\pm\sqrt{(\eps^a_\bp)^2+\Delta^2}\big)^2$. We have $\langle n_\bp|m_\bp\rangle=\delta_{nm}$ for $n,m=\pm$. The diagonalized Hamiltonian reads $\tilde\cH_\bp=U^{-1}_\bp\cH_\bp U_\bp$, which involves the complex eigenvalues 
\begin{align}
 \cE^\pm_\bp=\eps^s_\bp\pm\sqrt{(\eps^a_\bp)^2+\Delta^2}-i \Gamma_\bp \, .
\end{align}
The unitary transformation matrix is $U_\bp=\big(|+_\bp\rangle\,\,|-_\bp\rangle\big)$. Using the explicit form of the eigenstates in Eq.~\eqref{eqn:eigen_diag}, one can check that the Berry connection $i\langle n_\bp|\partial_{p_\alf}| n_\bp\rangle$ vanishes identically for both bands $n=\pm$ and all momenta.

The (diagonal) quasiparticle spectral function matrix is given by $\tilde A_\bp(\omega)=-\frac{1}{2\pi i}\big(\tilde\cG^R_\bp(\omega)-\tilde\cG^A_\bp(\omega)\big)$, which involves the (retarded) Green's function in the eigenbasis $\tilde \cG^R_\bp(\omega)=\big[\omega+\mu-\tilde \cH_\bp\big]^{-1}$ and $\tilde \cG^A_\bp(\omega)\equiv \big(\tilde \cG^R_\bp(\omega)\big)^\dag$. The diagonal components, the quasiparticle spectral functions of the upper and lower band $\tilde A^\pm_\bp(\omega)$, are Lorentzians at position $\re \cE^\pm_\bp-\mu=\eps^s_\bp\pm\sqrt{(\eps^a_\bp)^2+\Delta^2}-\mu$ and width $-\im \cE^\pm_\bp=\Gamma_\bp$. We calculate the spectral function matrix in the band basis $A_\bp(\omega)=-\frac{1}{2\pi i}\big(\cG^R_\bp(\omega)-\cG^A_\bp(\omega)\big)$ with $\cG^R_\bp(\omega)=[\omega+\mu-\cH_\bp]^{-1}$ and $\cG^A_\bp(\omega)\equiv \big(\cG^R_\bp(\omega)\big)^\dag$ with respect to the quasiparticle spectral functions $\tilde A^\pm_\bp(\omega)$. We decompose the quasiparticle Green's function of the upper and lower band into $\tilde \cG^{R,\pm}_\bp=\tilde P^\pm_\bp(\omega)-i\pi \tilde A^\pm_\bp(\omega)$ with $\tilde P^\pm_\bp(\omega)=\re \tilde G^\pm_\bp(\omega)$ and use $\cG^{}_\bp(\omega)=U^{}_\bp\tilde\cG^{}_\bp(\omega) U^{-1}_\bp$. Thus, the diagonal elements $A^\uparrow_\bp(\omega)$ and $A^\downarrow_\bp(\omega)$ 
of the spectral function matrix $A_\bp(\omega)$ read
\begin{alignat}{2}
 \label{eqn:Ahermitian}
 &A^{\uparrow/\downarrow}_\bp(\omega)&&=\frac{1}{2}\big(\tilde A^+_\bp+\tilde A^-_\bp\big)\pm\frac{1}{2}\frac{\eps^a_\bp}{\sqrt{(\eps^a_\bp)^2+\Delta^2}}\big(\tilde A^+_\bp-\tilde A^-_\bp\big)\, .
\end{alignat}
One can rewrite the prefactor in front of the quasiparticle spectral functions $\tilde A^\pm_\bp(\omega)$ by using the identity
\begin{align}
 \frac{1}{2}\bigg(1\pm\frac{\eps^a_\bp}{\sqrt{(\eps^a_\bp)^2+\Delta^2}}\bigg)=\frac{\Delta^2}{\Delta^2+(\eps_\bp-\re \cE^\pm_\bp)^2} \, .
\end{align}
The peaks of the quasiparticle spectral functions on the reconstructed Fermi surface defined by $\re \cE^\pm_\bp=\mu$ are thereby suppressed for momenta away from the bare Fermi surface (defined by $\eps_\bp=\mu$) \cite{Eberlein2016s}. The spectral function for single particle excitations $A_\bp(\omega) = A^\uparrow_{\bp-\bQ}(\omega) + A^\downarrow_\bp(\omega)$ thus resembles Fermi arcs, since only the pocket surface parts close to the bare Fermi surface are visible. Note that this result is independent of the momentum dependence of the relaxation rate $\Gamma_\bp$, which can lead to an additional reduction of the size of the spectral functions.


\section{Consequences of off-diagonal self-energy components}

In the main text, we considered only diagonal non-Hermitian contributions to the Hamiltonian. In the following, we discuss the consequences of non-Hermitian off-diagonal elements. The Hamiltonian $H_\bp$ combined with the most general non-Hermitian part reads
\begin{align}
\cH_\bp &= H_\bp-
i\begin{pmatrix}
  \Gamma^s_\bp+\Gamma^a_\bp & \Gamma^x_\bp+\Gamma^y_\bp \\[2mm]
  \Gamma^x_\bp-\Gamma^y_\bp & \Gamma^s_\bp-\Gamma^a_\bp
 \end{pmatrix} 
\end{align}
with real $\Gamma^x_\bp$ and $\Gamma^y_\bp$. The corresponding discriminant reads
\begin{align}
 \cD_\bp&=(\eps^a_\bp-i \Gamma^a_\bp)^2+(\Delta+i\Gamma^x_\bp)^2+(\Gamma^y_\bp)^2\\[2mm]&=(\eps^a_\bp)^2+\Delta^2-(\Gamma^a_\bp)^2-(\Gamma^x_\bp)^2+(\Gamma^y_\bp)^2-2i(\eps^a_\bp\Gamma^a_\bp-\Delta\,\Gamma^x_\bp) \, ,
\end{align}
whose zeros define the exceptional points. The diagonal part $\Gamma^s_\bp$ proportional to the identity matrix has no impact on the existence of exceptional points.  We assume that we are in the ordered state with $\Delta> 0$. In the case of $\Gamma^a_\bp=0$, a vanishing imaginary part of $\cD_\bp$ requires $\Gamma^x_\bp=0$. Thus, exceptional points do not exist since $(\eps^a_\bp)^2+\Delta^2+(\Gamma^y_\bp)^2\geq \Delta^2>0$. We see that a nonzero $\Gamma^a_\bp$ is necessary for exceptional points. For $\Gamma^a_\bp\neq 0$, the existence of exceptional points depends crucially on model and parameter details. We have checked that nonzero $\Gamma^x_\bp$ and $\Gamma^y_\bp$ have only an indirect impact on the quasiparticle dispersions $\cE^\pm_\bp$ and the spectral functions $\tilde A^\pm_\bp(\omega)$ and $A^{\uparrow/\downarrow}_\bp(\omega)$ by modifying the value of $\sqrt{\cD_\bp}$. Thus, we do not expect any major consequence on the conclusions that are presented in the main text.


\section{Behavior of the quasiparticle bands and eigenstates \\ when crossing the branch cut}

In the main text we have shown that the non-Hermitian Hamiltonian $\cH_\bp$ in Eq.~\eqref{eqn:DefH} can have exceptional points at isolated momenta. Exceptional points of opposite topological charge are connected by a one dimensional line, the branch cut, defined by the conditions $\eps^a_\bp=0$ and $(\Gamma^a_\bp)^2>\Delta^2$. We now discuss the behavior of the (complex) quasiparticle bands $\cE^\pm_\bp$ and the (biorthonormal) eigenstates when crossing the branch cut.


\subsection{Band exchange}

We analyze the discriminant $\cD_\bp=|\cD_\bp|\exp(i\arg \cD_\bp)$ and its (principle) square root $\sqrt{\cD_\bp} = \sqrt{|\cD_\bp|} \exp(i \arg\cD_\bp/2)$ on the nesting line defined by $\eps^a_\bp=0$, where $\arg$ is the argument function. We decompose the discriminant in Eq.~\eqref{eqn:disc} in its real and imaginary parts
\begin{align}
 \cD_\bp=\big(\eps^a_\bp-i\Gamma^a_\bp\big)^2+\Delta^2=(\eps^a_\bp)^2-(\Gamma^a_\bp)^2+\Delta^2-2i\eps^a_\bp\Gamma^a_\bp \, .
\end{align}
Thus, its absolute value and argument function read 
\begin{alignat}{2}
 &\lim_{\eps^a_\bp\rightarrow 0}|\cD_\bp|&&=|\Delta^2-(\Gamma^a_\bp)^2| \, ,\\[2mm]
 &\lim_{\eps^a_\bp\rightarrow 0^\pm}\arg\cD_\bp&&=\mp\pi\,\,\sign\big[\Gamma^a_\bp\big]\,\,\Theta\big[(\Gamma^a_\bp)^2-\Delta^2\big]0^+ \, ,
 \label{eqn:argD}
\end{alignat}
in the limit of vanishing $\eps^a_\bp$, where $\Theta(x)$ is the Heaviside step function and $0^+$ $(0^-)$ is an infinitesimal small positive (negative) number. Assuming that $\Gamma^a_\bp$ has no sign change when crossing the nesting line, the sign of $\arg\cD_\bp$ is determined by the sign of $\eps^a_\bp$, that is, it differs when approaching the nesting line from above zero, $\eps^a_\bp\rightarrow 0^+$ or $\eps^a_\bp\downarrow 0$, or from below zero, $\eps^a_\bp\rightarrow 0^-$ or $\eps^a_\bp\uparrow 0$. The discriminant is purely real on the nesting line. The square root of $\cD_\bp$ decomposes into a real and imaginary part $\sqrt{\cD_\bp}=\eps^\cD_\bp+i\Gamma^\cD_\bp$, which reads 
\begin{align}
 \label{eqn:epsD}
 &\lim_{\eps^a_\bp\rightarrow 0}\eps^\cD_\bp=\sqrt{|\Delta^2-(\Gamma^a_\bp)^2|}\,\,\Theta\big[\Delta^2-(\Gamma^a_\bp)^2\big]\, , \\[2mm]
 \label{eqn:GammaD}
 &\lim_{\eps^a_\bp\rightarrow 0^\pm}\Gamma^\cD_\bp=\mp\sign\big[\Gamma^a_\bp\big]\,\,\sqrt{|(\Gamma^a_\bp)^2-\Delta^2|}\,\,\Theta\big[(\Gamma^a_\bp)^2-\Delta^2\big] \, ,
\end{align}
when approaching the nesting line. Whereas the real part $\eps^\cD_\bp$ is continuous on the nesting line, the imaginary part $\Gamma^\cD_\bp$ is discontinuous on the branch cut, where $(\Gamma^a_\bp)^2>\Delta^2$. Thus, we find that the eigenvalues given in Eq.~\eqref{eqn:Epm}, $\cE^\pm_\bp=\eps^s_\bp\pm\eps^\cD_\bp-i(\Gamma^s_\bp\mp \Gamma^\cD_\bp)$ behave as
\begin{align}
 &\lim_{\eps^a_\bp\downarrow 0}\cE^\pm_\bp = \lim_{\eps^a_\bp\uparrow 0}\cE^\pm_\bp \hspace{0.5cm} \text{for} \hspace{0.5cm} |\Delta| > |\Gamma^a_\bp| \, , \\[2mm]
 &\lim_{\eps^a_\bp\downarrow 0}\cE^\pm_\bp=\lim_{\eps^a_\bp\uparrow 0}\cE^\mp_\bp \hspace{0.5cm} \text{for} \hspace{0.5cm} |\Delta| < |\Gamma^a_\bp| \, ,
\end{align}
when crossing the nesting line. On the branch cut, we have a band exchange \cite{Bergholtz2021s}.


\subsection{Simplifying gauge choice for biorthogonal eigenstates}

Before discussing the behavior of the eigenstates when crossing the branch cut, we summarize the gauge freedom for biorthogonal eigenstates.
The right and left eigenstates $|R^n_\bp\rangle$ and $|L^n_\bp\rangle$ of a (non-Hermitian) $N$-dimensional Hamiltonian $\cH_\bp$ are defined via the eigenvalue equations \cite{Brody2014s}
\begin{align}
 \label{eqn:eigenvalueEquation}
 \cH_\bp|R^n_\bp\rangle = \cE^n_\bp|R^n_\bp\rangle \, ,
 \hspace{2cm}
 \cH^\dag_\bp|L^n_\bp\rangle = (\cE^n_\bp)^*|L^n_\bp\rangle \, ,
\end{align}
respectively, where $\cE^n_\bp \in \mathds{C}$ are the non-degenerate (complex) eigenvalues of the bands $n = 1,\dots,N$ at momentum $\bp$. We further define $\langle R^n_\bp|=(|R^n_\bp\rangle)^\dag$ and $\langle L^n_\bp|=(|L^n_\bp\rangle)^\dag$. The left and right eigenstates are orthogonal with respect to the standard scalar product, that is
\begin{align}
 \label{eqn:norm}
 \langle L^n_\bp|R^m_\bp\rangle = (\langle R^m_\bp|L^n_\bp\rangle)^* = c^n_\bp\,\delta_{nm}
\end{align}
with $c^n_\bp \in \mathds{C}$. For $c^n_\bp = 1$ the matrix $\cU_\bp=(|R^1_\bp\rangle \dots |R^N_\bp\rangle)$ formed with the right eigenstates as columns, diagonalizes the Hamiltonian via $\cU^{-1}_\bp\cH^{}_\bp \cU^{}_\bp$, and its inverse matrix is $\cU^{-1}_\bp = (|L^1_\bp\rangle \dots |L^N_\bp\rangle)^\dag$.

The eigenvalue equations in Eq.~\eqref{eqn:eigenvalueEquation} determine the eigenstates up to a complex prefactor $r^n_\bp, l^n_\bp\in \mathds{C}$, so that
\begin{align}
 \label{eqn:scaling}
 |\tilde R^n_\bp\rangle=r^n_\bp|R^n_\bp\rangle \, ,
 \hspace{2cm}
 |\tilde L^n_\bp\rangle=l^n_\bp|L^n_\bp\rangle 
\end{align}
are eigenstates to the same eigenvalues $\cE^n_\bp$ and $(\cE^n_\bp)^*$, respectively. If we require that the scalar product in Eq.~\eqref{eqn:norm} remains independent of the rescaling gauge transformation in Eq.~\eqref{eqn:scaling}, we obtain the constraint
\begin{align}
 \label{eqn:normgauge}
 \langle L^n_\bp|R^m_\bp\rangle=\langle \tilde L^n_\bp|\tilde R^m_\bp\rangle \Leftrightarrow (l^n_\bp)^* r^n_\bp=1 \, .
\end{align}
Thus, the rescaling of the left eigenstate is determined by that of the right eigenstate \cite{Kozii2017s}. In order to construct a biorthonormal basis, $\langle \tilde L^n_\bp|\tilde R^n_\bp\rangle=1$, from a non-normalized basis satisfying Eq.~\eqref{eqn:norm}, we can rescale the left and right eigenvectors with $c^n_{L,\bp}$ and $c^n_{R,\bp}$ under the constraint $(c^n_{L,\bp})^* c^n_{R,\bp}=1/c^n_\bp$. Note that, in general, the prefactor for the normalization can be distributed differently between the left and right eigenstates. Further constraints will limit this freedom as discussed in the following. 

\paragraph{Hermitian Hamiltonian.}

For a Hermitian Hamiltonian $\cH^{}_\bp = \cH_\bp^\dag$, we see from Eq.~\eqref{eqn:eigenvalueEquation} that the left and right eigenstates get proportional to each other, $|L^n_\bp\rangle=f^n_\bp |R^n_\bp\rangle$ with $f^n_\bp\in \mathds{C}$. Requiring that the factor $f^n_\bp$ is unaffected by the rescaling gauge transformation defined in Eq.~\eqref{eqn:scaling}, the rescaling gauge has to be restricted to $l^n_\bp=r^n_\bp$. If we further require that the scalar product is independent of the gauge transformation (see Eq.~\eqref{eqn:normgauge}), we have $|l^n_\bp|=|r^n_\bp|=1$ and, thus, the well-known U(1) gauge freedom. If we require that the scalar products are independent of the distinction between left and right eigenstates, we find
\begin{align}
 \langle L^n_\bp|R^m_\bp\rangle=\langle R^n_\bp|R^m_\bp\rangle = \langle L^n_\bp|L^m_\bp\rangle \Leftrightarrow f^n_\bp=1 \, .
\end{align}
Thus, the left and right eigenstates have to be identical. This constrains the normalization to $c^n_{L,\bp}=c^n_{R,\bp}$.

\paragraph{Symmetric (non-Hermitian) Hamiltonian.}

In Eq.~\eqref{eqn:DefH}, we consider a non-Hermitian but symmetric Hamiltonian. For a symmetric Hamiltonian $\cH^{}_\bp=\cH^\text{tr}_\bp$, we see from Eq.~\eqref{eqn:eigenvalueEquation} that the complex conjugate of the left eigenstate and the right eigenstate are proportional,  
\begin{align}
 \label{eqn:constraintSym}
 \cH^{}_\bp=\cH_\bp^\text{tr}\Rightarrow (|L^n_\bp\rangle)^*=d^n_\bp|R^n_\bp\rangle
\end{align}
with $d^n_\bp\in \mathds{C}$. Requiring that the factor $d^n_\bp$ is unaffected by the rescaling gauge transformation in Eq.~\eqref{eqn:scaling}, the rescaling gauge has to be restricted to $(l^n_\bp)^*=r^n_\bp$. Requiring further that Eq.~\eqref{eqn:normgauge} is fulfilled, the rescaling gauge is restricted to $(r^n_\bp)^2=1$, that is, $r^n_\bp=l^n_\bp=\pm 1$. We only have a remaining $\mathds{Z}_2$ gauge freedom by choosing the sign at each momentum. Note that we chose and required a fixed $d^n_\bp$ for this conclusion. In case of a Hermitian symmetric Hamiltonian, we recover the U(1) gauge symmetry via relaxing the constraint on $d^n_\bp$.

\noindent For a biorthonormal basis with $c^n_\bp=1$, the Berry connection has the property 
\begin{align}
 \label{eqn:BerryConnection}
 \langle L^n_\bp|\partial^{}_{p_\alf} R^m_\bp\rangle =
 -\langle L^m_\bp|\partial^{}_{p_\alf} R^n_\bp\rangle -
 \delta^{}_{nm}\partial^{}_{p_\alf} \ln d^n_\bp \, .
\end{align}
We see that the diagonal contribution $n=m$ vanishes due to antisymmetry for a constant $d^n_\bp=d^n$. This simplifies the calculation of the geometric phase under parametric variation (here, momentum path in the Brillouin zone), since the geometric phase is entirely given by the overlap of the left eigenstate at the begin of the path and the right eigenstate at the end of the path \cite{Keck2003s}. Thus we choose $d^n_\bp=1$ in the following and in the main text. This choice restricts the freedom to choose the normalization to $(c^n_{L,\bp})^*=c^n_{R,\bp}$ and, thus, $c^n_{R,\bp}=1/\sqrt{c^n_\bp}$.


\subsection{Eigenstate sequence}

The right eigenstates of the upper and lower band $\cE^\pm_\bp=\eps^s_\bp\pm\sqrt{\cD_\bp}-i\Gamma^s_\bp$ read
\begin{align}
 \label{eqn:eigenstates}
 &|R^+_\bp\rangle=\frac{1}{\sqwp}
 \begin{pmatrix}
  \eps^a_\bp+\sqrt{\cD_\bp}-i\Gamma^a_\bp \\[1mm] -\Delta
 \end{pmatrix}\,, \hspace{1cm}
|R^-_\bp\rangle = \frac{1}{\sqwm}
\begin{pmatrix}
 \eps^a_\bp-\sqrt{\cD_\bp}-i\Gamma^a_\bp \\[1mm] -\Delta
\end{pmatrix}
\end{align}
with prefactor $w^\pm_\bp=\Delta^2+\big(\eps^a_\bp\pm\sqrt{\cD_\bp}-i\Gamma^a_\bp\big)^2$. In the limit $\Gamma^a_\bp=0$, we recover the eigenbasis given in Eq.~\eqref{eqn:eigen_diag}. Since $\cH_\bp$ is a symmetric matrix, the complex conjugated right eigenstate is proportional to the left eigenstate. We choose $|L^\pm_\bp\rangle=(|R^\pm_\bp\rangle)^*$. By the proper choice of $w^\pm_\bp$ the left and right eigenstates form a biorthonormal basis, $\langle L^n_\bp|R^m_\bp\rangle=\delta_{nm}$. The remaining gauge freedom in Eq.~\eqref{eqn:scaling} that conserves the relation between the left and right eigenstates and the orthonormality is $|\tilde R^\pm_\bp\rangle=\pm|R^\pm_\bp\rangle$. According to Eq.~\eqref{eqn:BerryConnection}, the Berry connection vanishes, that is $\langle L^\pm_\bp|\partial^{}_{p_\alf} R^\pm_\bp\rangle=0$, which can be easily verified by the explicit from in Eq.~\eqref{eqn:eigenstates}.

In the following, we discuss the behavior of the eigenstates in Eq.~\eqref{eqn:eigenstates} when crossing the nesting line ($\eps^a_\bp=0$) and, in particular, the branch cut ($(\Gamma^a_\bp)^2>\Delta^2$), which reduces to the behavior of $\sqrt{\cD_\bp}$ and the normalization constants $\sqwp$ and $\sqwm$. We discussed the behavior of the real and imaginary parts of $\sqrt{\cD_\bp}=\eps^\cD_\bp+i\Gamma^\cD_\bp$ in Eqs.~\eqref{eqn:epsD} and \eqref{eqn:GammaD}, respectively. For a different choice of the normalization, $|\tilde R^n_\bp\rangle=\sq{w^n_\bp}\,|R^n_\bp\rangle$ and $|\tilde L^n_\bp\rangle=1/\big(\sq{w^n_\bp}\big)^*\,|L^n_\bp\rangle$, we immediately see that $|\tilde R^\pm_\bp\rangle\rightarrow |\tilde R^\pm_\bp\rangle$ for $\Delta^2>(\Gamma^a_\bp)^2$ and $|\tilde R^\pm_\bp\rangle\rightarrow |\tilde R^\mp_\bp\rangle$ for $(\Gamma^a_\bp)^2>\Delta^2$ when crossing the nesting line. The same holds for the left eigenstates $|\tilde L^n_\bp\rangle$. Thus, the eigenstates $|\tilde R^n_\bp\rangle$ are exchanged when crossing the branch cut similarly to the exchange of the eigenvalues \cite{Bergholtz2021s}. Note that the eigenstate $|\tilde R^n_\bp\rangle=\sq{w^n_\bp}\,|R^n_\bp\rangle$, which does not fulfill the momentum-independent relation between the left and right eigenstates but instead $(|\tilde L^n_\bp\rangle)^*=d^n_\bp |\tilde R^n_\bp\rangle$ with $d^n_\bp=1/w^n_\bp$ in Eq.~\eqref{eqn:constraintSym}, has a non-vanishing Berry connection
\begin{align}
 \langle \tilde L^n_\bp|\partial^{}_{p_\alf} \tilde R^m_\bp\rangle = \langle L^n_\bp|\frac{1}{\sq{w^n_\bp}}\partial^{}_{p_\alf}\big[\sqrt{\vphantom{\cD}\smash{w^m_\bp}}|R^m_\bp\rangle\big]=\frac{1}{2}\delta^{}_{nm}\,\partial^{}_{p_\alf}\ln w^n_\bp\,,
\end{align}
where we used $\langle L^n_\bp|R^m_\bp\rangle=\delta_{nm}$ and $\langle L^\pm_\bp|\partial^{}_{p_\alf} R^\pm_\bp\rangle=0$. Thus, both the direct overlap and the Berry curvature contribute to the geometric phase for $|\tilde R^n_\bp\rangle$ in contrast to $|R^n_\bp\rangle$, where only the direct overlap contribute. 

Based on the previous arguments, we see that the choice of the normalization in Eq.~\eqref{eqn:eigenstates} simplifies the analysis of the behavior of the eigenstates under parametric variation. We expect that the behavior of $\sqwp$ and $\sqwm$ is crucial when crossing the branch cut. Thus, we continue by analyzing
\begin{align}
 w^\pm_\bp&=\Delta^2+\big(\eps^a_\bp\pm\eps^\cD_\bp-i(\Gamma^a_\bp\mp\Gamma^\cD_\bp)\big)^2
 \\[2mm]
 &= \Delta^2+\big(\eps^a_\bp\pm\eps^\cD_\bp\big)^2-\big(\Gamma^a_\bp\mp\Gamma^\cD_\bp\big)^2 -2i\big(\eps^a_\bp\pm\eps^\cD_\bp\big)\big(\Gamma^a_\bp\mp\Gamma^\cD_\bp\big) \, ,
\end{align}
when approaching the nesting line with $\eps^a_\bp=0$. For $\Delta^2-(\Gamma^a_\bp)^2>0$, we use Eqs.~\eqref{eqn:epsD} and \eqref{eqn:GammaD} and obtain
\begin{align}
 &\lim_{\eps^a_\bp\rightarrow 0}\re w^\pm_\bp=2\big(\Delta^2-(\Gamma^a_\bp)^2\big)\, , \\
 &\lim_{\eps^a_\bp\rightarrow 0}\im w^\pm_\bp=\mp 2\Gamma^a_\bp\sqrt{\Delta^2-(\Gamma^a_\bp)^2} \, .
\end{align}
These results are equal when approaching the nesting line from above or from below zero, $\lim_{\eps^a_\bp\downarrow 0}w^\pm_\bp=\lim_{\eps^a_\bp\uparrow 0}w^\pm_\bp$. We conclude
\begin{align}
 \lim_{\eps^a_\bp\downarrow 0}|R^\pm_\bp\rangle=\lim_{\eps^a_\bp\uparrow 0}|R^\pm_\bp\rangle \hspace{1cm}\text{for}\hspace{1cm} \Delta^2>(\Gamma^a_\bp)^2\, .
\end{align}
The crossing behavior through the branch cut defined by $(\Gamma^a_\bp)^2>\Delta^2$ is more intriguing. The imaginary part of $w^\pm_\bp$ vanishes on the branch cut. We have to distinguish the sign of the imaginary part when approaching zero, which will be crucial in the crossing behavior of $\sq{w^\pm_\bp}$. Using the relation $\arg \cD_\bp = -i \ln\cD_\bp/|\cD_\bp|$ and the identities $\eps^\cD_\bp=\sqrt{|\cD_\bp|}\cos\hspace{-0.8mm}\big(\hspace{-0.6mm}\arg\cD_\bp/2\big)$ and $\Gamma^\cD_\bp=\sqrt{|\cD_\bp|} \sin\hspace{-0.8mm}\big(\hspace{-0.6mm}\arg\cD_\bp/2\big)$, we obtain the expansion
\begin{alignat}{2}
 &\eps^\cD_\bp&&\rightarrow \frac{|\Gamma^a_\bp\eps^a_\bp|}{\sqrt{(\Gamma^a_\bp)^2-\Delta^2}}+\mathcal{O}\big[(\eps^a_\bp)^2\big]\,,\\[1mm]&\Gamma^\cD_\bp&&\rightarrow -\sign\big[\Gamma^a_\bp\eps^a_\bp\big]\,\sqrt{(\Gamma^a_\bp)^2-\Delta^2}+\mathcal{O}\big[(\eps^a_\bp)^2\big] \, .
\end{alignat}
Thus, the limit of the real part of $w^\pm_\bp$ when approaching the nesting line reads
\begin{align}
 \lim_{\eps^a_\bp\rightarrow 0}\re w^\pm_\bp=-2\big((\Gamma^a_\bp)^2-\Delta^2\big)\bigg[1\pm \frac{\sign[\eps^a_\bp]}{\sqrt{1-(\Delta/\Gamma^a_\bp)^2}}\bigg] \, .
\end{align}
Note that the absolute value of the last term in the square bracket is truly larger one since $0<(\Delta/\Gamma^a_\bp)^2<1$ on the branch cut in the ordered phase with $\Delta>0$. Thus, it modifies the overall sign depending on the band $n=\pm$ and the sign of $\eps^a_\bp$. The limit of the imaginary part of $w^\pm_\bp$ when approaching the nesting line reads
\begin{align}
 \lim_{\eps^a_\bp\rightarrow 0}\im w^\pm_\bp
 &=\lim_{\eps^a_\bp\rightarrow 0}\bigg(-2\eps^a_\bp\Gamma^a_\bp\bigg[1\pm\frac{\sign[\eps^a_\bp]}{\sqrt{1-(\Delta/\Gamma^a_\bp)^2}}\bigg]\bigg[1\pm \sign[\eps^a_\bp]\sqrt{1-(\Delta/\Gamma^a_\bp)^2}\bigg]\bigg)\\[0mm] &= 
 \mp \sign[\Gamma^a_\bp]\,0^+ \, .
\end{align}
We denote the approach of zero from positive values as $0^+$. In the second step, we used that the last square bracket is always positive on the branch cut. Note that the sign only depends on the band $n=\pm$ and the sign of $\Gamma^a_\bp$, but not on the sign of $\eps^a_\bp$. We can read off the argument of $w^\pm_\bp$ and obtain
\begin{align}
 &\lim_{\eps^a_\bp\downarrow 0}\arg w^+_\bp=-\,\sign[\Gamma^a_\bp]\,\,\pi\,, && \lim_{\eps^a_\bp\uparrow 0}\arg w^+_\bp=0\,,\\
 &\lim_{\eps^a_\bp\downarrow 0}\arg w^-_\bp=0\,, && \lim_{\eps^a_\bp\uparrow 0}\arg w^-_\bp=+\,\sign[\Gamma^a_\bp]\,\,\pi\,.
\end{align}
Using the definition of the principle square root
\begin{align}
 \sqrt{w^\pm_\bp}=\sqrt{|w^\pm_\bp|}e^{i\arg w^\pm_\bp/2}
\end{align}
we obtain the limiting behavior
\begin{align}
 &\lim_{\eps^a_\bp\downarrow 0} \sqr{w^+_\bp}=-\lim_{\eps^a_\bp\uparrow 0}\sqr{w^-_\bp}\, \hspace{1cm}\text{and}\hspace{1cm}\lim_{\eps^a_\bp\uparrow 0}\sqr{w^+_\bp}=\lim_{\eps^a_\bp\downarrow 0}\sqr{w^-_\bp} \hspace{1cm}\text{for}\hspace{1cm} (\Gamma^a_\bp)^2>\Delta^2\,.
\end{align}
Note the sign change depends on the sign of $\eps^a_\bp$ when approaching the nesting line. There is no dependence on the sign of $\Gamma^a_\bp$, since $\exp(\pm i\pi)=-1$. This is in contrast to the topological invariant in Eq.~\eqref{eqn:vorticity} due to the behavior of $\arg\cD_\bp$ given in Eq.~\eqref{eqn:argD} at the branch cut. Combining the results leads to
\begin{align}
 \lim_{\eps^a_\bp\downarrow 0}|R^+_\bp\rangle =
 -\lim_{\eps^a_\bp\uparrow 0}|R^-_\bp\rangle \, , \quad
 \lim_{\eps^a_\bp\downarrow 0}|R^-_\bp\rangle =
  \lim_{\eps^a_\bp\uparrow 0}|R^+_\bp\rangle \, ,
\end{align}
corresponding to Eq.~\eqref{eqn:Evolution} in the main text.


\subsection{Evolution of the eigenstates and the quasiparticle velocity through the crossing point}

The previous results do not depend on the precise path in the Brillouin zone through the nesting line, even if the branch cut is crossed. In the main text, we considered a path along the Fermi surface, defined by the constraint $\re \cE^\pm_\bp=\mu$. In presence of exceptional points and, thus, branch cuts, hole and electron pockets can merge at a single point, which allows for, in principle, six paths through the crossing point along the Fermi surface.
A concrete case is shown in Fig.~\ref{fig:Crossing}. We now discuss the evolution of the (right) eigenstates $|n_\bp\rangle\equiv |R^n_\bp\rangle$ on the six possible paths. In addition, we show the quasiparticle velocity $\partial_{p_\alf}E^n_\bp$ in $\alf=x,y$ direction. 

\begin{figure}[t!]
\centering
    \includegraphics[width=0.3\textwidth]{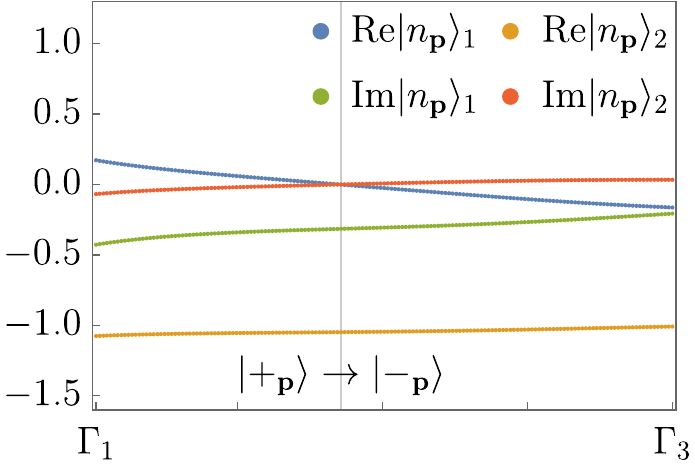}
    \includegraphics[width=0.3\textwidth]{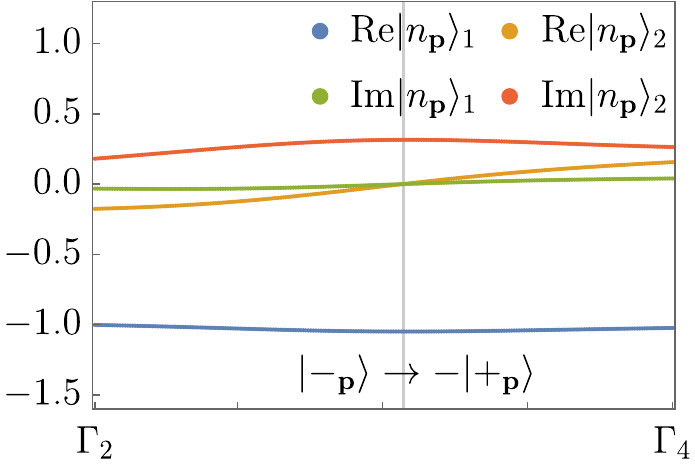} \\[3mm]
   \includegraphics[width=0.3\textwidth]{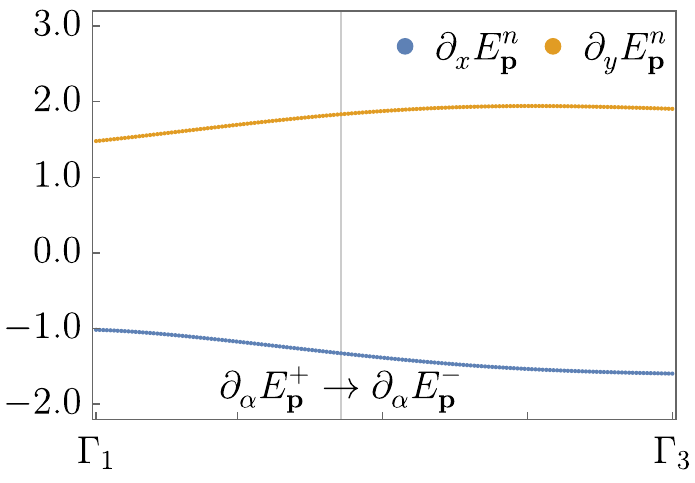}
   \includegraphics[width=0.3\textwidth]{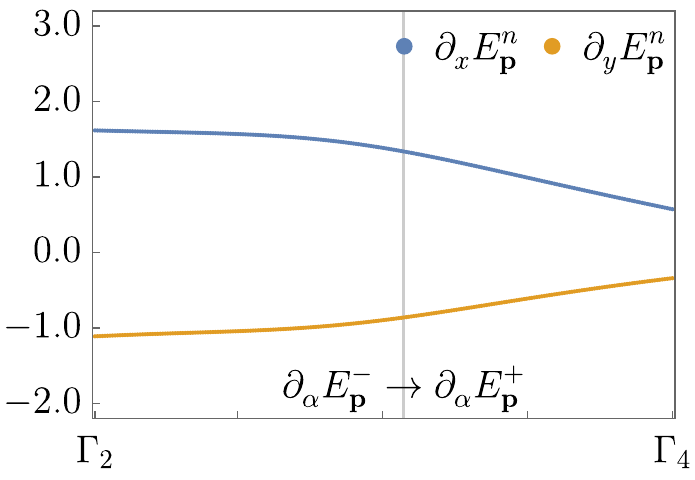}
\caption{The components of the eigenstates $|n_\bp\rangle\equiv |R^n_\bp\rangle$ in Eq.~\eqref{eqn:eigenstates} (top row) and the quasiparticle velocity (bottom row) on the two diagonal paths $\Gamma_1\rightarrow \Gamma_3$ (left column) and $\Gamma_2\rightarrow \Gamma_4$ (right column) with starting and endpoints as indicated in Fig.~\ref{fig:Crossing}. The vertical line indicates the crossing point at which the eigenstates and bands are exchanged. Note the sign change in the eigenstates. The eigenstates and the quasiparticle velocities are continuous at the crossing point. \label{fig:Crossing1}}
\end{figure}
In Fig.~\ref{fig:Crossing1}, we show the components of $|n_\bp\rangle$ and the quasiparticle velocity on the two diagonal paths $\Gamma_1 \rightarrow \Gamma_3$ and $\Gamma_2 \rightarrow\Gamma_4$. The vertical line indicates the momentum at which the hole and electron pockets merge. The respective changes of the eigenstates and bands are indicated. On both paths the branch cut is crossed with a sign change of $\eps^a_\bp$ from minus to plus (white to gray in Fig.~\ref{fig:Crossing}). According to the general result in Eq.~\eqref{eqn:Evolution}, there is a transition $|+_\bp\rangle \to |-_\bp\rangle$ at the crossing point between $\Gamma_1$ and $\Gamma_3$, and a transition $|-_\bp\rangle \to - |+_\bp\rangle$ at the crossing point between $\Gamma_2$ and $\Gamma_4$. One can see that the eigenstates and the quasiparticle velocities are continuous at the crossing point, in spite of the branch cut. Hence, these are the paths followed by the electrons in semiclassical dynamics.

\begin{figure}[t!]
\centering
   \includegraphics[width=0.24\textwidth]{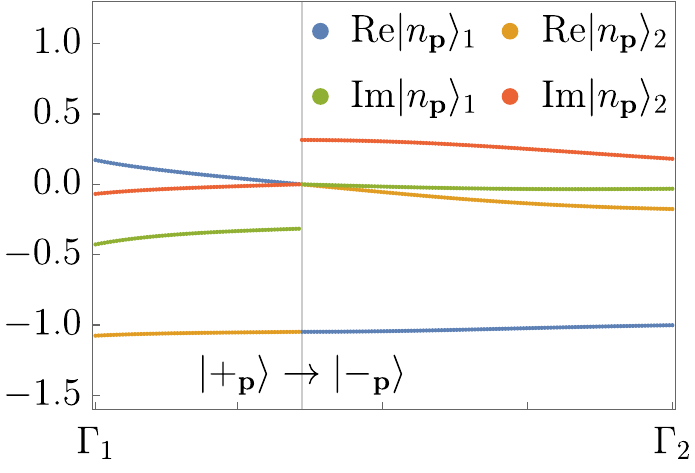}
   \includegraphics[width=0.24\textwidth]{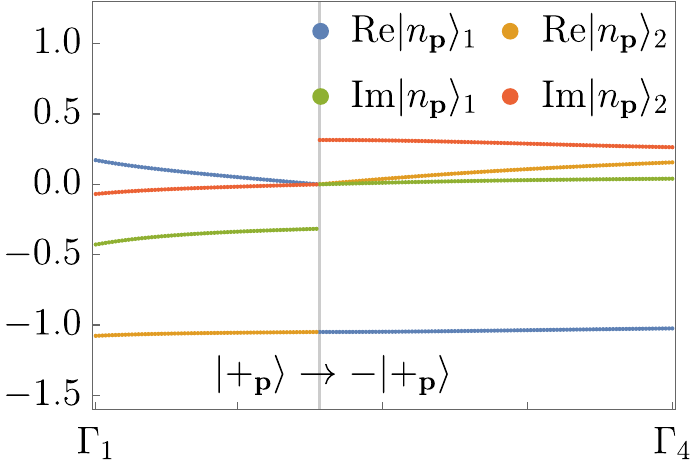}
   \includegraphics[width=0.24\textwidth]{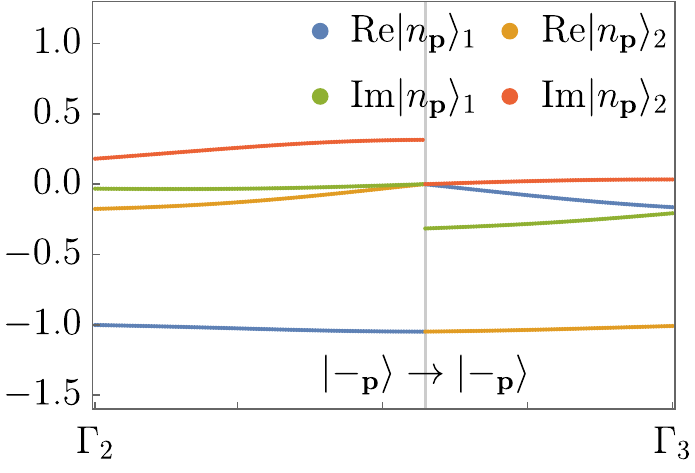}
   \includegraphics[width=0.24\textwidth]{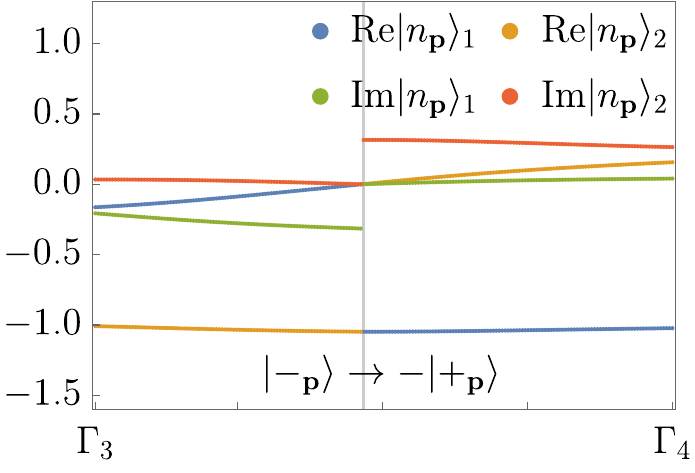}\\[3mm]
   \includegraphics[width=0.24\textwidth]{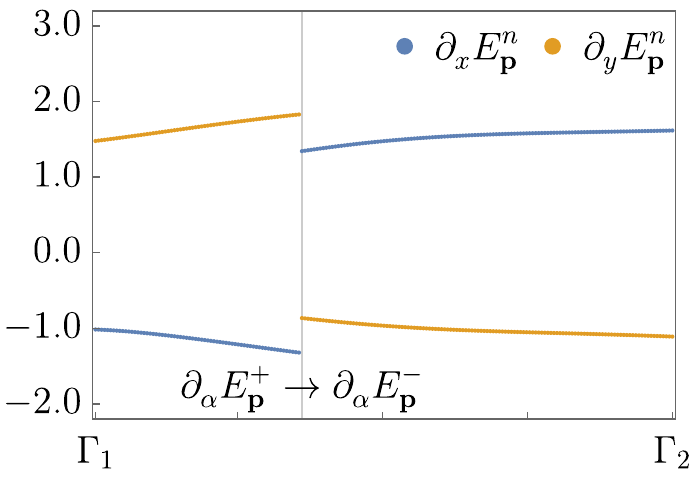}
   \includegraphics[width=0.24\textwidth]{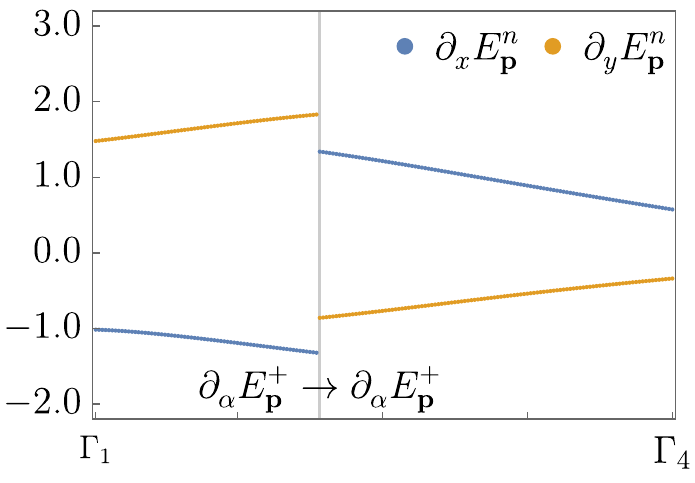}
   \includegraphics[width=0.24\textwidth]{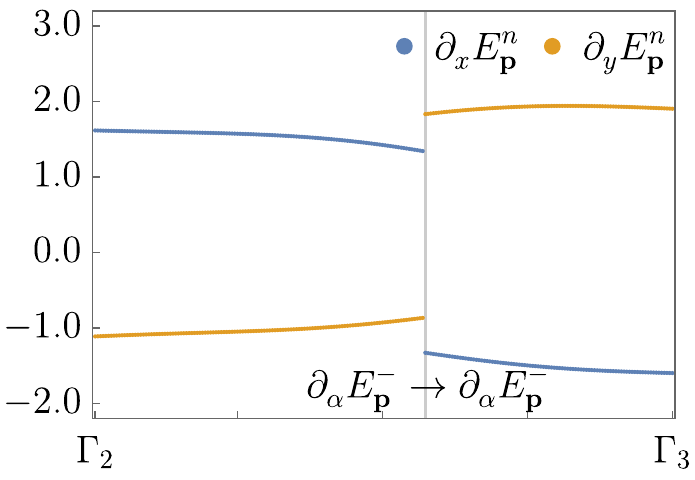}
   \includegraphics[width=0.24\textwidth]{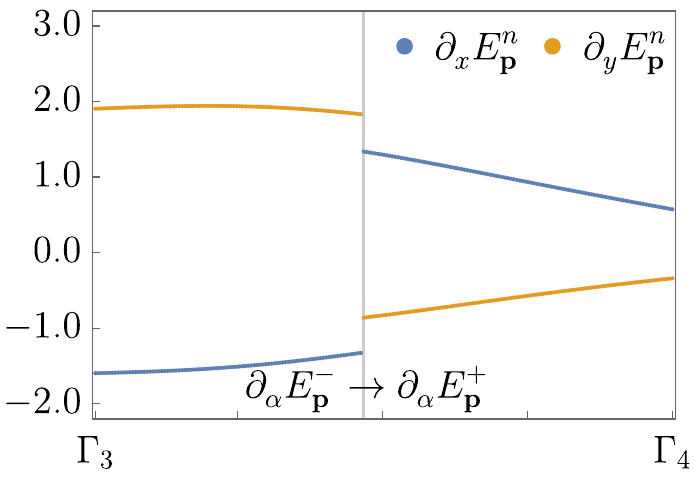}
\caption{The components of the eigenstates $|n_\bp\rangle\equiv |R^n_\bp\rangle$ in Eq.~\eqref{eqn:eigenstates} (top row) and the quasiparticle velocity (bottom row) on the non-diagonal paths with starting and endpoints as indicated in Fig.~\ref{fig:Crossing}. The vertical line indicates the crossing point at which the eigenstates and bands are exchanged. The eigenstates and the quasiparticle velocities exhibit discontinuities at the crossing point.
\label{fig:Crossing2}}
\end{figure}
In Fig.~\ref{fig:Crossing2}, we show the components of $|n_\bp\rangle$ and the quasiparticle velocity on the four non-diagonal paths through the crossing point. For these paths discontinuities are present at the crossing point.


\section{Derivation and properties of the spectral functions}

We calculate the spectral function in the bare band basis in terms of the quasiparticle quantities $\tilde \cP_\bp(\omega)=\re \tilde \cG^R_\bp(\omega)$ and $\tilde \cA_\bp(\omega)=-\frac{1}{\pi}\im \tilde \cG^R_\bp(\omega)$. The spectral function in the bare band basis reads
\begin{align}
 \cA_\bp(\omega)&=-\frac{1}{2\pi i}\Big(\cG^R_\bp(\omega)-\cG^A_\bp(\omega)\Big)\\&=-\frac{1}{2\pi i}\Big(\cU^{}_\bp\tilde\cG^R_\bp(\omega)\cU^{-1}_\bp-\big[\cU^{}_\bp\tilde\cG^R_\bp(\omega)\cU^{-1}_\bp\big]^\dag\Big) \, ,
\end{align}
where $\cU_\bp$ diagonalizes the Hamiltonian $\cH_\bp$ in Eq.~\eqref{eqn:DefH}. We use the decomposition $\tilde \cG^R_\bp(\omega)=\tilde \cP_\bp(\omega)-i\pi\tilde\cA_\bp(\omega)$. Evaluating the two diagonal elements $A^\uparrow_\bp(\omega)$ and $A^\downarrow_\bp(\omega)$ of $\cA_\bp(\omega)$ leads to the result in Eqs.~\eqref{eqn:A1}-\eqref{eqn:A3}.

For small $\Gamma^a_\bp$, we have 
\begin{align}
 \eps^\cD_\bp=\sqrt{(\eps^a_\bp)^2+\Delta^2}+\mathcal{O}\big[(\Gamma^a_\bp)^2\big]\,, \hspace{2cm}
 \Gamma^\cD_\bp=-\frac{\eps^a_\bp\Gamma^a_\bp}{\sqrt{(\eps^a_\bp)^2+\Delta^2}}+\mathcal{O}\big[(\Gamma^a_\bp)^2\big] \, ,
\end{align}
and, thus, up to first order in $\Gamma^a_\bp$, the spectral function
\begin{align}
 A^{\uparrow/\downarrow}_\bp(\omega) &= \frac{1}{2}\big(\tilde A^+_\bp(\omega)+\tilde A^-_\bp(\omega)\big)\pm\frac{1}{2}\frac{\eps^a_\bp}{\sqrt{(\eps^a_\bp)^2+\Delta^2}}\big(\tilde A^+_\bp(\omega)-\tilde A^-_\bp(\omega)\big)\\&\pm\frac{1}{2\pi}\frac{\Delta^2}{(\eps^a_\bp)^2+\Delta^2}\frac{\Gamma^a_\bp}{\sqrt{(\eps^a_\bp)^2+\Delta^2}}\big(\tilde P^+_\bp(\omega)-\tilde P^-_\bp(\omega)\big) \, .
\end{align}
The formula of $A^{\uparrow/\downarrow}_\bp(\omega)$ reduces to \eqref{eqn:Ahermitian} in the limit $\Gamma^a_\bp=0$. We see that a nonzero $\Gamma^a_\bp$ naturally leads to the unconventional contribution in the second row that involves the quasiparticle principle function $\tilde P^\pm_\bp(\omega)$, which is the real part of the retarded quasiparticle Green's function. 

\begin{figure}[t!]
\centering
   \includegraphics[width=0.35\textwidth]{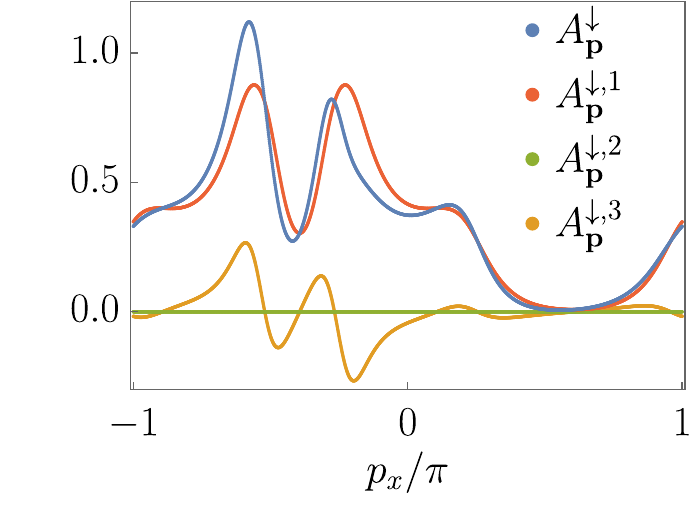}
   \includegraphics[width=0.35\textwidth]{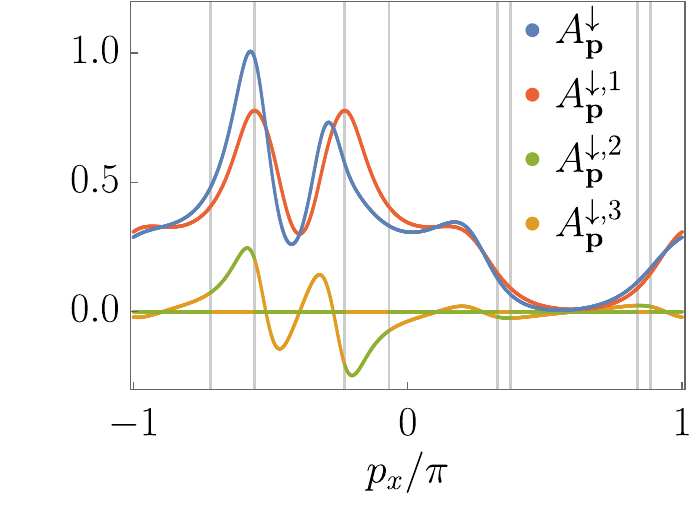}
\caption{The spectral function $A^\downarrow_\bp(0)$ and its three contributions given in Eqs~\eqref{eqn:A1}-\eqref{eqn:A3} along the upper the nesting line ($\eps^a_\bp=0$, $p_y > 0$) for $\gamma_d/t=0.8$ (left) and $\gamma_d/t=1$ (right). The crossings through the exceptional points are indicated by gray vertical lines.\label{fig::Cont}}
\end{figure}

For general $\Gamma^a_\bp$, we will see now that this unconventional contribution in Eq.~\eqref{eqn:A3} regularizes the contribution in Eq.~\eqref{eqn:A2} at the exceptional point, where $\cD_\bp=0$ and the Hamiltonian is not diagonalizable. We express the complex discriminant in polar form 
\begin{align}
 \cD^{}_\bp=r^\cD_\bp\big(\cos\varphi^\cD_\bp+i\sin\varphi^\cD_\bp\big) \,
\end{align}
where $r^\cD_\bp=|\cD_\bp|\geq 0$ and $\varphi^\cD_\bp=\arg \cD_\bp\in (-\pi,\pi]$. At the exceptional point we have $r^\cD_\bp=0$. The complex phase $\varphi^\cD_\bp$ has a $\pi$-jump when crossing the exceptional point. We use $\eps^\cD_\bp=\sqrt{r^\cD_\bp}\cos(\frac{1}{2}\varphi^\cD_\bp)$ and $\Gamma^\cD_\bp=\sqrt{r^\cD_\bp}\sin(\frac{1}{2}\varphi^\cD_\bp)$ and expand up to first order in $\sqrt{r^\cD_\bp}$. The first term of $A^{\uparrow/\downarrow}_\bp(\omega)$ in Eq.~\eqref{eqn:A1} reads 
\begin{align}
 \frac{1}{2}\big(\tilde A^+_\bp(\omega)+\tilde A^-_\bp(\omega)\big)=\frac{1}{\pi}\frac{\Gamma^a_\bp}{(\Gamma^a_\bp)^2+(\omega+\mu-\eps^s_\bp)^2} + \mathcal{O}\Big[\big(\sqrt{r^\cD_\bp}\big)^2\Big]\, .
\end{align}
It is independent of $r^\cD_\bp$ and $\varphi^\cD_\bp$ and, thus, continuous when crossing the exceptional point from any direction. The second and third term in Eq.~\eqref{eqn:A2} and Eq.~\eqref{eqn:A3} involves the difference of the quasiparticle spectral functions and the principle functions with a particular prefactor $c_i$, respectively, which we denote as $\eqref{eqn:A2}+\eqref{eqn:A3}\equiv c_2 f_2+c_3 f_3$. The $f_i$ read up to first order in $\sqrt{r^\cD_\bp}$ 
\begin{alignat}{2}
  &f_2&&\equiv\frac{1}{2}\big(\tilde A^+_\bp(\omega)-\tilde A^-_\bp(\omega)\big)\\[2mm]& &&=\frac{1}{\pi}\frac{2\Gamma^a_\bp\big(\omega+\mu-\eps^s_\bp\big)\cos(\frac{1}{2}\varphi^\cD_\bp)-\big(\omega+\mu-\eps^s_\bp-\Gamma^a_\bp\big)\big(\omega+\mu-\eps^s_\bp+\Gamma^s_\bp\big)\sin(\frac{1}{2}\varphi^\cD_\bp)}{\big((\Gamma^s_\bp)^2+(\omega+\mu-\eps^s_\bp)^2\big)^2}\,\,\sqrt{r^\cD_\bp} 
 + \mathcal{O}\Big[\big(\sqrt{r^\cD_\bp}\big)^3\Big]\, ,\\[3mm]
 &f_3&&\equiv\frac{1}{2\pi}\big(\tilde P^+_\bp(\omega)-\tilde P^-_\bp(\omega)\big)\\[2mm]& &&=\frac{1}{\pi}\frac{2\Gamma^s_\bp\big(\omega+\mu-\eps^s_\bp\big)\sin(\frac{1}{2}\varphi^\cD_\bp)+\big(\omega+\mu-\eps^s_\bp-\Gamma^s_\bp\big)\big(\omega+\mu-\eps^s_\bp+\Gamma^s_\bp\big)\cos(\frac{1}{2}\varphi^\cD_\bp)}{\big((\Gamma^s_\bp)^2+(\omega+\mu-\eps^s_\bp)^2\big)^2}\sqrt{r^\cD_\bp} + \mathcal{O}\Big[\big(\sqrt{r^\cD_\bp}\big)^3\Big]\, .
\end{alignat}
As expected, $f_2$ and $f_3$ vanish in the limit $\sqrt{r^\cD_\bp}\rightarrow 0$ since the two eigenstate states $n=\pm$ are identical at the exceptional point. The two prefactors read
\begin{alignat}{2}
 &c_2=\frac{\eps^a_\bp \eps^\cD_\bp-\Gamma^a_\bp\Gamma^\cD_\bp}{(\eps^\cD_\bp)^2+(\Gamma^\cD_\bp)^2}&&=\frac{\eps^a_\bp\cos(\frac{1}{2}\varphi^\cD_\bp)-\Gamma^a_\bp\sin(\frac{1}{2}\varphi^\cD_\bp)}{\sqrt{r^\cD_\bp}}\,,\\[2mm]
 &c_3=\frac{\eps^a_\bp \Gamma^\cD_\bp+\Gamma^a_\bp\eps^\cD_\bp}{(\eps^\cD_\bp)^2+(\Gamma^\cD_\bp)^2}&&=\frac{\eps^a_\bp\sin(\frac{1}{2}\varphi^\cD_\bp)+\Gamma^a_\bp\cos(\frac{1}{2}\varphi^\cD_\bp)}{\sqrt{r^\cD_\bp}} \,.
\end{alignat}
Thus, in general, the combinations $c_2f_2$ and $c_3f_3$ are finite in the limit $\sqrt{r^\cD_\bp}\rightarrow 0$. Note that on the nesting line with $\eps^a_\bp=0$, we have $\varphi^\cD_\bp=0$ and, thus, $c_2=0$ for $\Delta^2_\bp>(\Gamma^a_\bp)^2$ and $\varphi^\cD_\bp=\pi$ and, thus, $c_3=0$ for $(\Gamma^a_\bp)^2>\Delta^2_\bp$. We see that the terms in Eq.~\eqref{eqn:A2} and and in Eq.~\eqref{eqn:A3} are both discontinuous at the exceptional point. However, due to the phase discontinuity of $\varphi^\cD_\bp$ when crossing the exceptional points, we have
\begin{align}
 c_2f_2+c_3f_3\rightarrow (\mp c_3)(\mp f_3)+(\pm c_2)(\pm f_2)=c_2f_2+c_3f_3 \hspace{1cm}\text{for}\hspace{1cm} \varphi^\cD_\bp\rightarrow \varphi^\cD_\bp\pm\pi \, , 
\end{align}
that is, the second and third term of $A^{\uparrow/\downarrow}_\bp(\omega)$ are mapped onto each other at the exceptional point, so that the sum is continuous. In Fig.~\ref{fig::Cont}, we show the spectral function $A^\downarrow_\bp(0)$ at zero frequency for momenta on the upper nesting line (see Fig.~\ref{fig:FermiSurface}) and its three contributions $A^{\downarrow,i}_\bp(0)$, which are defined by Eqs.~\eqref{eqn:A1}-\eqref{eqn:A3}. For $\gamma_d/t=0.8$ (left), no exceptional points are present. We have $\Delta^2>(\Gamma^2_\bp)^2$ and, thus, $A^{\uparrow,2}_\bp$ vanishes. The unconventional contribution $A^{\uparrow,3}_\bp$ is finite for almost all momenta. For $\gamma_\bp/t=1$ (right), the nesting line crosses the eight exceptional points (gray vertical lines), at which $A^{\downarrow,2}_\bp$ and $A^{\downarrow,3}_\bp$ are discontinuous. However, the sum is continuous and does not show any feature at the exceptional points.


\end{document}